\documentclass[12pt,preprint]{aastex}
\def\h1{$^1$H}
\def\o16{$^{16}$O}
\def\c12{$^{12}$C}
\def\n14{$^{14}$N}

\def\he4{$^4$He}
\def\sun{\ M$_{\odot}$}
\def\mkms{\rm km s^{-1}}
\def\lya{Ly$\alpha$}

\newcommand{\nhi}{$N_{\rm HI}$}
\everymath{\rm}
\everydisplay{\sf}
\received{}
\revised{}
\accepted{}
\shortauthors{Nava \& Henry}
\shorttitle{N and O At Low Metallicity}
\begin{document}
\title{THE CHEMICAL EVOLUTION OF HIGH Z GALAXIES FROM THE RELATIVE ABUNDANCES OF N, Si, S, AND Fe IN DAMPED LYMAN ALPHA SYSTEMS} 

\author {R.B.C. Henry}

\affil{The Homer L. Dodge Department of Physics \& Astronomy, University of Oklahoma,
Norman, OK 73019; henry@nhn.ou.edu}

\and

\author {Jason X. Prochaska}

\affil{Lick Observatory, University of California, 1156 High Street, Santa Cruz, 
CA 95064; xavier@ucolick.org}

\begin{abstract}

Abundances of N, Si, S, and Fe for 45 damped Lyman alpha systems (DLAs) have been compiled and detailed one-zone chemical evolution models have been constructed for 30 of them. Assuming continuous star formation, we found that final abundances in each object can be modelled by adjusting only two parameters, i.e. its time-averaged star formation efficiency and evolutionary age, with ranges in our sample of 0.01-1.5~Gyr$^{-1}$ and 0.18-2.0~Gyr, respectively. In addition, average star formation efficiency and evolutionary age appear to be anticorrelated for the sample, suggesting that the star formation efficiency in a typical DLA decreases with age. At the same time, N/Si in DLAs is directly linked to an object's age. There is an apparent bimodality in the distribution of N/Si values which could be the result of a statistical accident or an effect produced by a truncated or flattened IMF. We find that the mean and small dispersion of Si/Fe values is related to the generally young ages of DLAs, wherein not all Fe has yet been expelled by Type Ia supernovae.  Finally, the large scatter and generally lower values of N/Si of DLAs with respect to blue compact galaxies, despite their partially overlapping metallicities, indicate that DLAs are generally younger than the latter.

\end{abstract}



\section{INTRODUCTION}

Damped \lya\ systems (DLAs) serve as valuable probes of heavy element abundances and chemical evolution at high redshifts and thus large lookback times (Lu et al. 1996; Pettini, Lipman, \& Hunstead 1995). They provide complementary information to that obtained from stellar and nebular studies of nearby galaxies (Pritzl, Venn, \& Irwin 2005; Bresolin, Garnett, \& Kennicutt 2004).

The canonical view of DLAs is that they are galactic systems, perhaps representing a variety of morphological types, whose interstellar matter absorbs light from QSOs positioned behind them along the observer's line of sight. These systems have H~I surface densities where $N(H~I) \ge 2.0 \times 10^{20}$cm$^{-2}$ (roughly 2~M$_{\odot}$pc$^{-2}$), i.e. high enough so that absorption occurs in both the core as well as the damping wings of the \lya\ feature. Under such optical depth conditions metal lines become detectable, and careful analysis of their profiles enables researchers to measure abundances of more than 20 elements in some DLAs
(e.g.\ Prochaska, Howk, \& Wolfe 2003). 
For a comprehensive discussion of numerous aspects of DLAs, including their contributions to the cosmic abundance picture, the reader is referred to the recent review by Wolfe, Gawiser, \& Prochaska (2005).

Two approaches to understanding DLA abundances exist in the literature. The ``global'' one treats abundance averages over co-moving volumes and attempts to model the observations using global values of star formation histories (Edmunds \& Phillipps 1997). 
In contrast, the ``local''approach studies individual DLAs one at a time, using standard chemical evolution equations to interpret observations for a sample of objects (Dessauges-Zavadsky et al. 2006, 2007). 
We focus on the local method here.

Confronting abundance measurements in DLAs with numerical models has proved fruitful for understanding the chemical evolution of these objects. Using this approach, Matteucci, Molaro, \& Vladilo (1997) found that abundance patterns in DLAs are consistent with bursting star formation in a dwarf irregular galaxy, thus implicating the latter as progenitors for some DLAs. Later, a similar analysis by Calura, Matteucci, \& Vladilo (2003) appeared to rule out elliptical galaxies as possible progenitors but could not decide between spiral disks, irregular galaxies, or bursting dwarf irregulars for the host. Dessauges-Zavadsky et al. (2004) employed both a spiral disk and dwarf irregular galaxy model in efforts to explain abundance measurements in three objects and concluded that DLAs are either outer regions of spiral disks which have undergone slow continuous star formation or are dwarf irregular galaxies which have experienced either continuous or bursting star formation. Their models had the added virtue of predicting star formation rates which were consistent with the rates inferred by Wolfe, Prochaska, \& Gawiser (2003)
from observed \ion{C}{2}$^*$ absorption. 
The recent paper by Dessauges-Zavadsky et al. (2006) used models from Dessauges-Zavadsky et al. (2004) but extended their survey to report abundances of 22 elements for a sample of 11 objects. Finally, Dessauges-Zavadsky et al. (2007) has computed chemical evolution models of six additional DLAs using the same approach as they did in the 2004 paper and reached similar conclusions regarding the plausible star formation histories for these objects.

Metallicity in DLAs [M/H], as traced by [Si/H], [S/H], and [Zn/H]\footnote{We adopt the standard bracket notation, $[X/H] \equiv log(X/H)-log(X/H)_{\odot}$}, ranges from roughly $-0.5$ to $-2.6$\,dex , with a cosmic mean metallicity  [M/H]$\approx -1.1$\,dex 
(Prochaska \& Wolfe 2000; Prochaska et al.\ 2003; Pettini 2004).   
The observations indicate the
metallicity increases with the age of the universe and that 
no DLA has a metallicity extending below solar by more 
than -2.6~dex (Prochaska et al.\ 2003).
These conclusions have been bourne out by
more recent compilations 
(Meiring et al.\ 2006; Ledoux et al.\ 2006; Prochaska et al.\ 2007).
The typical values of the alpha abundance, as traced by Si and S, are 1-2 orders of magnitude below solar values.

DLA abundance studies over the past decade have focused much of their attention on two important element/element ratios, i.e. N/$\alpha$ and $\alpha$/Fe, where the relevant alpha element is usually Si, with O and S sometimes playing the role. 
The choice of Si is not ideal given that this element is mildly refractory,
but the resonance-line \ion{O}{1} transitions are generally saturated and the
triplet of \ion{S}{2} transitions near $\lambda \approx 1250$\AA\ are
frequently lost in the \lya\ forest.

The characteristic range of the N/$\alpha$ ratio, as well as the manner in which objects appear to form a bimodal distribution, has sparked much debate. 
These issues are of particular interest in determining overall N production in galaxies, because
the evolution of this ratio is linked to the relative roles of both low 
and intermediate mass stars (LIMS) on the one hand and massive stars on the other.   
Pettini et al. (2002) employed VLT UVES observations along with published data to study the N/O ratio in 10 DLAs and confirmed the broad range in values seen earlier by Lu et al. (1998). Pettini et al. proposed that this dispersion could be explained if LIMS produced a large fraction of the N, and that their sample objects represented various stages of N release by LIMS, which evolve somewhat slower than massive stars. This proposal was consistent with chemical evolution models of Henry, Edmunds, \& K{\"o}ppen (2000; hereafter HEK). Then Prochaska et al. (2002) reported 
a possible bimodality in the distribution of N/$\alpha$ later supported
by Centurion et al.\ (2003). 
They found that while most DLAs show $[N/\alpha] \approx -0.5$, 
a small group of objects (dubbed ``low nitrogen DLAs'' or LNDLAs) 
have $[N/\alpha] \approx -1.5$. 
They considered several explanations for this bimodality, including the delay in N release favored by Pettini et al. and concluded that this would not likely result in a bimodality, claiming that under this scenario the development of a potential gap between the normal N objects and the LNDLAs would likely not occur. Instead, their preferred hypothesis was a reduction in the effective N yield effected by a highly flattened or truncated IMF possessing a deficit of LIMS. (Prochaska et al. assumed that LIMS were responsible for most of the N production in the Universe, an issue which is yet to be settled.)

Following the claim of an N/$\alpha$ distribution bimodality, Molaro (2003) and D'Odorico \& Molaro (2004) confirmed the bimodality and argued in favor of the delay hypothesis. Lanfranchi \& Matteucci (2003) showed with detailed models that the top-heavy IMF proposal of Prochaska et al. (2002) could indeed explain the low N/$\alpha$ values in LNDLAs,  but at the expense of significantly underproducing Fe, with the result that the predicted Si/Fe ratio is much larger than observed. 
More recently, Chiappini, Matteucci, \& Meynet (2003) and Chiappini, Matteucci, \& Ballero (2005) have employed the stellar yields of Meynet \& Maeder (2002) to show that the LNDLAs could arise in systems undergoing short bursts (10-200~Myr), while the star formation history of most DLAs is characterized by significantly longer bursts (800~Myr).  
Finally, the N/$\alpha$ problem was addressed in the work of Moll{\'a} et al. (2006), who showed that by using the yields for LIMS computed by Buell (1997) and subsequently published and used in Gavil{\'a}n, Buell, \& Moll{\'a} (2005) and Gavil{\'a}n, Moll{\'a}, \& Buell (2006), they could explain a large range in N/$\alpha$ in astronomical objects, including the high values recently found in halo stars by Spite et al. (2005) and Israelian et al. (2004), the moderate values seen in blue compact galaxies, and normal as well as low nitrogen DLAs. Clearly, the problem of N/$\alpha$ is far from resolved.

Another major problem concerning DLAs is the attempt to explain the supersolar $\alpha$/Fe ratios observed in nearly all DLAs. For example, in their extensive study of DLA abundances Prochaska \& Wolfe (2002) determined a mean value of log(Si/Fe) of +0.3~dex with a small standard deviation. This finding is consistent with the recent results of Dessauges-Zavadsky et al. (2006) who derived a mean value of log(Si/Fe) of +0.430~dex with a standard deviation of 0.114~dex.  
The majority of authors (e.g.\ Vladilo 1999; Pettini et al.\ 2000; 
Centurion et al.\ 2000; Nissen et al.\ 2004) have argued that 
the observed supersolar ratios of Si/Fe are simply the result of depletion; if the Fe abundance is corrected for depletion the Si/Fe becomes solar. 
However, Wolfe, Gawiser, \& Prochaska (2005)
plot [Si/Fe] versus [Si/H] and argue that in the limit of vanishing metallicity, where dust depletion is insignificant, [Si/Fe] approaches $\approx$ +0.3, not zero. 
Furthermore, Dessauges-Zavadsky et al.\ (2006) have fitted the Si/Fe ratio
as a function of observed Zn/Fe value (a measure of differential dust 
depletion) and found supersolar Si/Fe ratios for ``dust-free'' systems.
Resolution of the Si/Fe ambiguity is of crucial importance for understanding the chemical nature of DLAs; 
a true solar value implies that these systems are much more advanced 
chemically than if Si/Fe is significantly above solar.

{\it The goal of this paper is to explore the N/$\alpha$ and $\alpha$/Fe problems just discussed for DLAs, as well as other questions related to their metallicity range, by employing a very basic chemical evolution model with few {\rm a priori} assumptions along with the smallest number of variable parameters.} To this end we have compiled published abundances of N, Si, S, and Fe for a sample of 45 DLAs. Using only those objects for which firm values of N and Si abundances are available (that is, objects with only limits for one or both of these elements were rejected), we computed a detailed chemical evolution model for each object, where the only parameters which were varied were the star formation efficiency\footnote{The star formation efficiency (SFE) is the mass of new stars formed per mass of material available for forming stars per unit time. The unit for the SFE is therefore Gyr$^{-1}$.} (SFE) and evolutionary age (measured from the time when star formation began to the observed epoch.). We then used the results to address the abundance problems presented above. 

Section~2 presents and discusses the abundance compilation upon which our analysis is based, while \S~3 describes the implications of the data. The results and implications of the chemical evolution models are provided in \S~4, and our conclusions are presented in \S~5. 

\section{THE DATABASE}

Table \ref{tab:stat} provides a summary of all of the DLAs with N constraints
in the literature. 
Columns 1 and 2 contain the name of the background quasar and redshift of the absorbing system, while columns 3-10 give abundance information. The literature sources for the abundances are given in the last column. 

All DLAs were observed with either the HIRES
(Vogt et al. 1994) or UVES (Dekker et al. 2000) echelle spectrometers on the
Keck 10m and VLT 8m telescopes, respectively.
The data have high resolution (FWHM~$< 10 \mkms$) and moderate
to high signal-to-noise (S/N~$>10 {\rm pix^{-1}}$). 
The sample includes $\approx 10$ new observations acquired with
the HIRES spectrometer using the upgraded mosaic CCD (Prochaska et al.\ 2007, in prep).
These data have been reduced and analyzed with the same techniques
described in Prochaska et al. (2002).

The nitrogen and $\alpha$ elemental abundances were derived
from low-ion transitions and reflect gas-phase abundances.
The ionic column densities were measured with line-profile fits
or the apparent optical depth method (Savage \& Sembach 1991).
The reported errors only indicate statistical uncertainty.
One may allow for an additional $\approx 0.05$\,dex error for systematic
uncertainty, e.g.\ continuum placement.
This is especially true for the N\,I transitions,
all of which lie within the \lya\ forest.

In general, one must consider ionization corrections and 
differential depletion in order to convert the gas-phase ionic column densities
to intrinsic elemental abundances.  Conveniently, N, S, and O 
are non-refractory and Si appears to be only 
mildly depleted (Savage \& Sembach 1996; Welty et al.\ 2006); 
[Si/S]~$\approx 0$ in those DLA where both are measured (Prochaska \& Wolfe 2002; Dessauges-Zavadsky et al. 2006).
Ionization corrections, however, may be important for nitrogen, which has
a relatively large cross-section to X-rays (Sofia \& Jenkins 1998).
Approximately half of the DLAs listed in Table~\ref{tab:stat}
have had their ionization states assessed in detail, and several DLAs have
been corrected by a few 0.1\,dex, while others have been removed from further
analysis because of large uncertainty (Prochaska et al. 2002).
We suspect that a few of the other DLAs drawn from the literature,
e.g. Centuri{\'o}n et al. (2003), may require modest ionization corrections
which would increase the N abundance by 0.1 to 0.3\,dex.
While these systematic corrections exceed the reported statistical
uncertainty, we emphasize that effects of this order have little
bearing on our conclusions.

\begin{figure}
\centering
\includegraphics[width=12cm,angle=270]{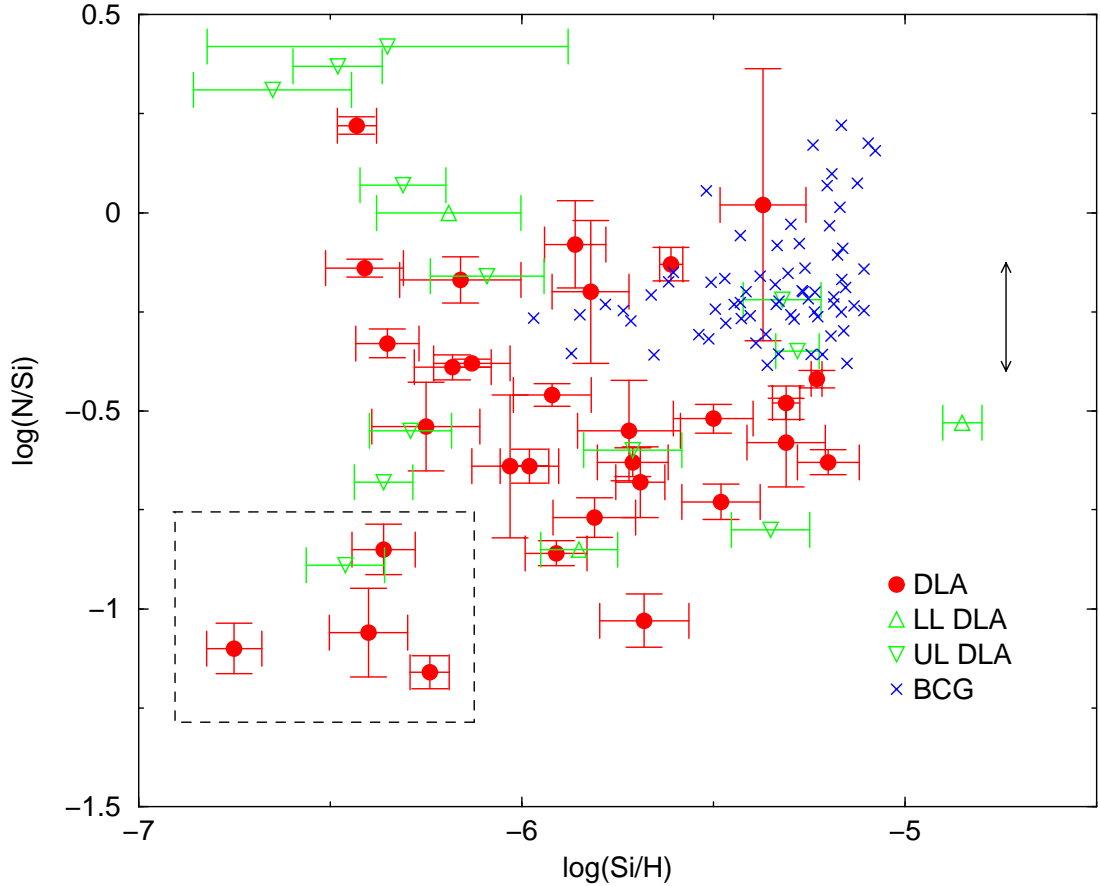}
\caption{log(N/Si) versus log(Si/H) showing the positions and uncertainties for DLAs in our sample with firm abundance measurements (filled circles), as well as those objects with upper (down triangles) or lower (up triangles) limits. Also included are the measurements for a sample of blue compact galaxies from Nava et al. (2006) in which their O abundances have been transformed into Si abundances using the O/Si solar abundance ratio of Asplund et al. (2005). The double-arrowed vertical line indicates the extent of the N/O plateau for BCGs as defined by Nava et al. The box in the lower left region surrounds the low nitrogen DLAs.}
\label{n2sivsi}
\end{figure}

Observed values of log(N/Si) and log(Si/H) from Table~\ref{tab:stat} are plotted against each other in Fig.~\ref{n2sivsi}. The filled circles show 30 points for DLAs with firm values, while up/down open triangles indicate an additional 15 objects where only lower/upper limits for log(N/Si) could be determined. For comparison purposes we also show positions for 68 blue compact galaxies (BCG) taken from Nava et al.'s (2006) nitrogen study. In order for the BCGs to be placed on this plot the O abundances given by Nava et al. were converted to Si abundances using log(Si/O)$_{\odot}$ = -1.15 
(Asplund et al. 2005). We remind the reader that each object represented in Fig.~\ref{n2sivsi} has evolved independently from the others and has a unique star formation history. Each point indicates an object's current position in N/Si--Si/H space.

It is interesting to note the relatively large range in log(N/Si) exhibited by DLAs compared with BCGs. Nava et al. defined an N/O plateau as comprising those objects forming a narrow horizontal track whose vertical dimension is indicated with a double-headed arrow in Fig.~\ref{n2sivsi}. We note that all but a few DLAs with firm values for N/Si lie on or below this plateau, with Q1008+36 appearing to have the smallest N/Si ratio, -1.15, of all the DLAs in Table~\ref{tab:stat}. 
In contrast, the DLA with the highest value of log(N/Si), +0.22, is J1558-0031, a system with relatively simple kinematics
that also shows D absorption (O'Meara et al.\ 2006). 
This vertical range of -1.15 to +0.22 for DLAs can be compared 
with the range in N/O for the plateau of BCGs given by Nava et al. (and converted to N/Si) of -0.39 to -0.12. The log of the mean N/Si values for 
DLAs is -0.40 (-0.46 when J1558-0031 is excluded), compared to -0.28 for BCGs. 
Restricting the BCG sample to the $-0.39$ to $-0.12$ interval and the
DLA sample to log(Si/H)~$>-6$, a two-sided Kolmogorov-Smirnov test rules
out the null hypothesis that the N/Si distributions of the two samples
are drawn from the same parent population.
Finally, Fig.~\ref{n2sivsi} shows that DLA metallicities, as measured 
here by Si, extend below the lowest BCG metallicity by as much as 0.6~dex. 

The LNDLAs discussed in the Introduction are visible as the group of four objects with firm abundances (filled circles) and one upper limit (open triangle) located in the lower left of Fig.~\ref{n2sivsi} inside the box. Focusing on the circles for a moment, there is an apparent void stretching obliquely from upper left to lower right that separates these objects from the bulk of the DLA population. 

\begin{figure}
\centering
\includegraphics[width=12cm,angle=270]{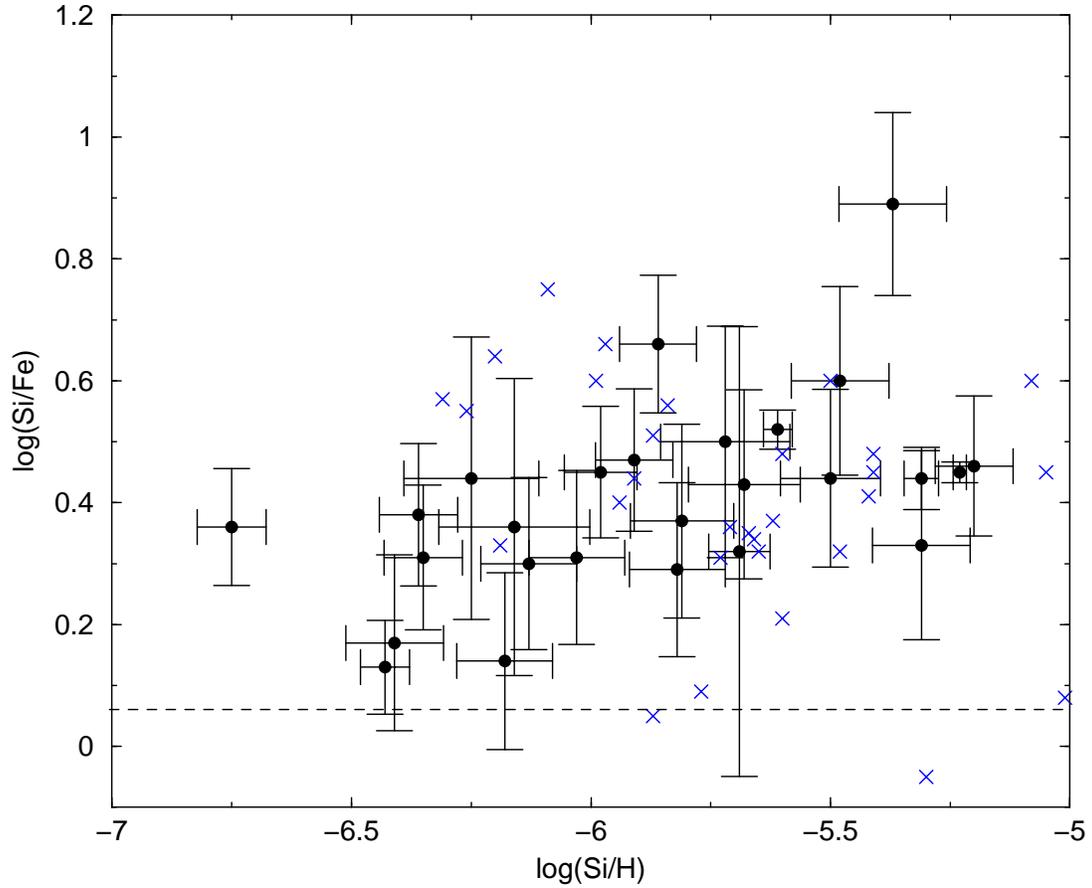}
\caption{log(Si/Fe) versus log(Si/H) showing the positions and uncertainties for DLAs in our sample.  Globular cluster data from Pritzl et al. (2005) are indicated with crosses. The solar value for log(Si/H) is -4.49, outside the range of the plot. The horizontal dashed line shows the solar abundance ratio for log(Si/Fe), where the solar values were taken from Asplund et al. (2005).}
\label{si2fevsi}
\end{figure}

A second interesting pair of quantities to consider is log(Si/Fe) versus log(Si/H), as the first parameter illustrates the $\alpha$/Fe characteristics of DLAs. Only objects in Table~\ref{tab:stat} with firmly measured abundances of Si and Fe are shown in Fig.~\ref{si2fevsi}. The solar value for log(Si/Fe) from Asplund et al (2005) is +0.06 and is indicated with a horizontal dashed line. The vertical range in log(Si/Fe) extends from +0.13 to +0.89, while the log of the mean of Si/Fe is +0.44.  With a solar value for log(Si/H) -4.49, the DLAs fall roughly between -0.71 and -2.3~dex below solar metallicity. Our figure should be compared with a similar one in the bottom panel of Fig.~8 in Wolfe, Gawiser, \& Prochaska (2005) in which it is suggested that Si/Fe is constant at low Si/H but turns upward when [Si/H]$\ge$-1. The general consensus is
that the increase in Si/Fe above one-tenth solar metallicity is due to increasing depletion of Fe onto dust as the metallicity rises (Prochaska \& Wolfe 2002).

This general pattern is very similar to what is observed for stars of low metallicity in the Milky Way Galaxy and dwarf spheroidals. Recently, Pritzl, Venn, \& Irwin (2005) have compiled numerous abundances for Galactic globular cluster stars. Data for their objects which fall within the Si/H range of our plot are shown as crosses in Fig.~\ref{si2fevsi}. Keeping in mind the likely differences in the star formation histories of DLAs and the Milky Way, we nevertheless find that the log of their mean value for Si/Fe for all of their objects is +0.44, identical to our DLA value.
Supersolar values of Si/Fe are thought to arise in stars when they form from material which has been enriched primarily by massive star production of alpha elements such as Si, while type Ia supernovae have yet to expel completely their Fe peak products.

\begin{figure}
\centering
\includegraphics[width=12cm,angle=270]{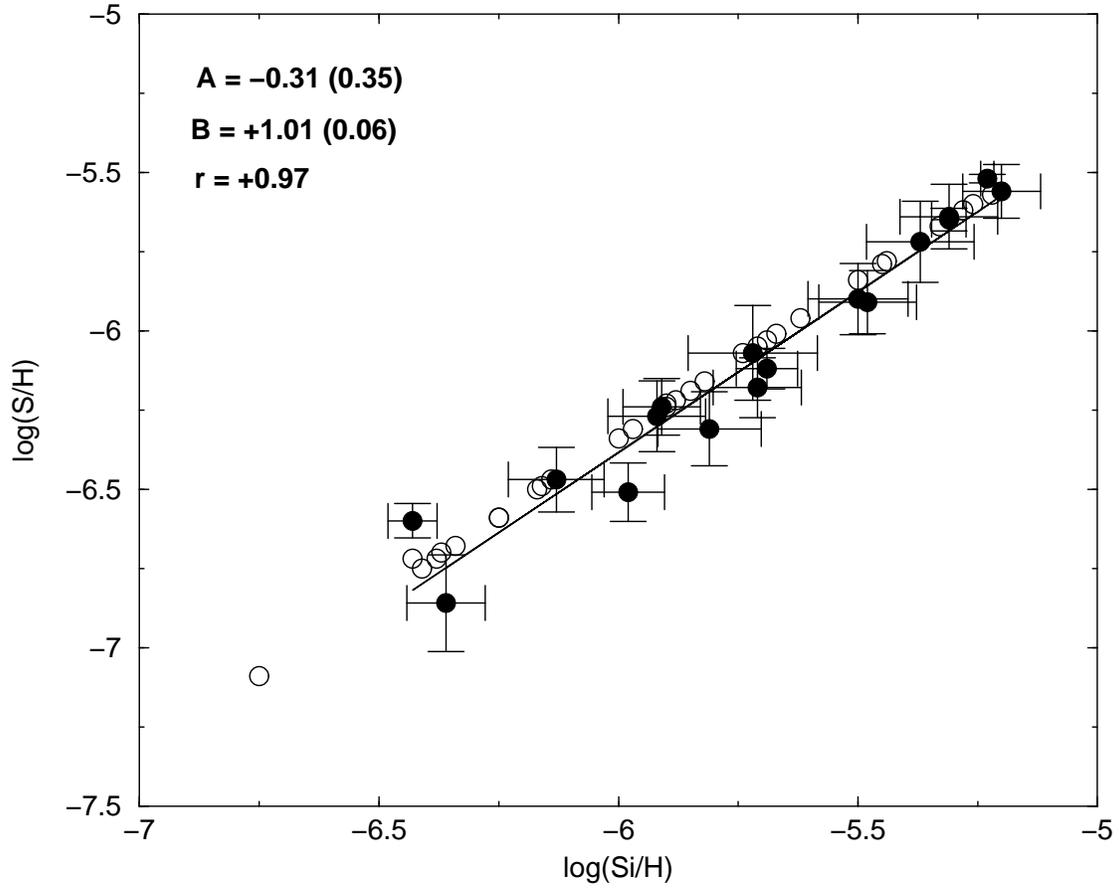}
\caption{log(S/H) versus log(Si/H) showing observed values and uncertainties for our sample objects as filled circles. Model-predicted values for the same objects are shown as open circles. The bold line shows the results of a least squares fit to the observations, with values (and uncertainties) for the y-intercept (A), slope (B), and correlation coefficient (r) given in the upper left corner.}
\label{svsi}
\end{figure}

Finally, in Fig.~\ref{svsi} we present a plot of log(S/H) versus log(Si/H) for the 19 objects in Table~\ref{tab:stat} with firm values for both of these elements. The bold line shows a least squares fit to the data, and the elements of the fit are given in the upper left corner of the graph. The correlation coefficient (r) and slope (B) are both near unity, signifying a solid linear relation between these two elements. The value for A, -0.31, is nearly equal to the solar value for log(S/Si) of -0.37 given by Asplund et al. (2005).
 
In summary, the following points can be made about the patterns seen in the data:

\begin{enumerate}

\item The N/Si ratios exhibit a large amount of scatter for DLAs compared with the relatively narrow distribution in N/O observed in blue compact galaxies.

\item N/Si and Si/H in DLAs generally fall below corresponding values for blue compact galaxies.

\item The distribution of DLAs in the N/Si--Si/H plane appears to be bimodal (see \S\ 4.2.1). 

\item Observed values of Si/Fe for DLAs are consistently above the solar level for the same ratio.

\item A tight correlation exists between S/H and Si/H.

\end{enumerate}

We now endeavor to explore the above characteristics of DLAs using chemical evolution models.
In doing so we focus only on those objects for which firm values for both N and Si abundances are available; objects for which only limits have been determined for either N or Si are ignored. This culling reduced our sample to 30 objects. 

\section{ANALYSIS}

An individual chemical evolution model for each of the 30 sample objects was calculated, after which model predictions of N/Si, Si/Fe, Si/H, S/H, and Fe/H were compared with their observed counterparts. Details of the chemical evolution code that we employed are provided in Appendix~A. Each model was a one-zone calculation,
that assumed continuous star formation, the presence of infall (but not outflow), a stellar mass range between 0.1 and 120\sun, a Salpeter (1955) IMF, no instantaneous recycling, instantaneous mixing, and a standard set of stellar yields. This is our {\it basic model}. The only quantities which varied from model to model were the input SFE and evolutionary age. The abundance levels of each of the isotopes $^4$He, $^{12}$C, $^{14}$N, $^{16}$O, $^{20}$Ne, $^{28}$Si, $^{32}$S, $^{35}$Cl, $^{36}$Ar, and $^{56}$Fe were tracked over time. 

For our entire analysis, massive star yields were taken from Portinari et al. (1998) for all of the above isotopes but $^{35}$Cl and $^{36}$Ar; yields for these last two species came from Woosley \& Weaver (1995). Yields of LIMS for $^4$He, $^{12}$C, $^{14}$N came from Marigo (2001). Neither yield set accounts for stellar rotation. Contributions of SNIa to the evolution of all but $^4$He were included using the yields for the W7 and W70 models of Nomoto et al. (1997). All carbon yields were multiplied by 0.3 in keeping with our findings in Henry, Nava, \& Prochaska (2006). 

Table \ref{singlemodels} shows the results of the model of each DLA in our culled sample of 30 objects. Object names are provided in column 1, while columns~2 and 3 list the SFE and age used as input in the model. Columns~4-12 list the observed and model predicted values for the quantity specified in the column heading. To facilitate comparisons, each table element is given in the format {\it observed/model}. Ellipses are entered when no observational values are available. Since few DLA carbon abundances have been determined, only model predictions are given in columns 9 and 12. The last two rows of the table contain the average observed uncertainty relevant to each column and the solar abundances from Asplund et al. (2005), respectively.  

During the modeling process, the principal goal was to adjust the SFE and age in order to reproduce specifically the observed values of N/Si and Si/H. Therefore, observation and theory are in particularly close agreement for these two ratios (columns~10 and 5, respectively). Sulfur and iron predictions, while not primary modeling targets, are nevertheless in good agreement with the observations. 
Major discrepancies occur in the case of oxygen, however, where typically the models overpredict the abundance by an average of 0.43 dex.
We speculate that part of this difference may be due to line saturation
in the observed \ion{O}{1} transitions, causing observed O/H to be underestimated.
  
Only a small number of our objects have C and O abundances available in the literature. DLA carbon abundances in particular are extremely difficult to measure because the only resonance line outside of the \lya\ forest, C~II $\lambda$1334.5, becomes saturated even in metal-poor systems.
We find that the ratio of C/Si is roughly the same throughout our DLA sample, with an average value in our models of +0.22 ($\sigma$=.059).
D'Odorico \& Molaro (2004) observed a value of +0.66 dex for C/Si in BR1202-07. 
Centuri{\'o}n et al. (2003) report O abundances for three of the objects in our sample, and these observations are included in Table~\ref{singlemodels}. We predict an O abundance only 0.04 dex larger than the observed value for Q0000-2619, although our predictions for Q0347-38 and Q2206-19 are 0.23 dex and 0.33 dex less than measured levels, respectively. We also found that O/H scales tightly with Si/H in our models, as expected from massive star nucleosynthesis.

\begin{figure}
\centering
\includegraphics[width=12cm,angle=270]{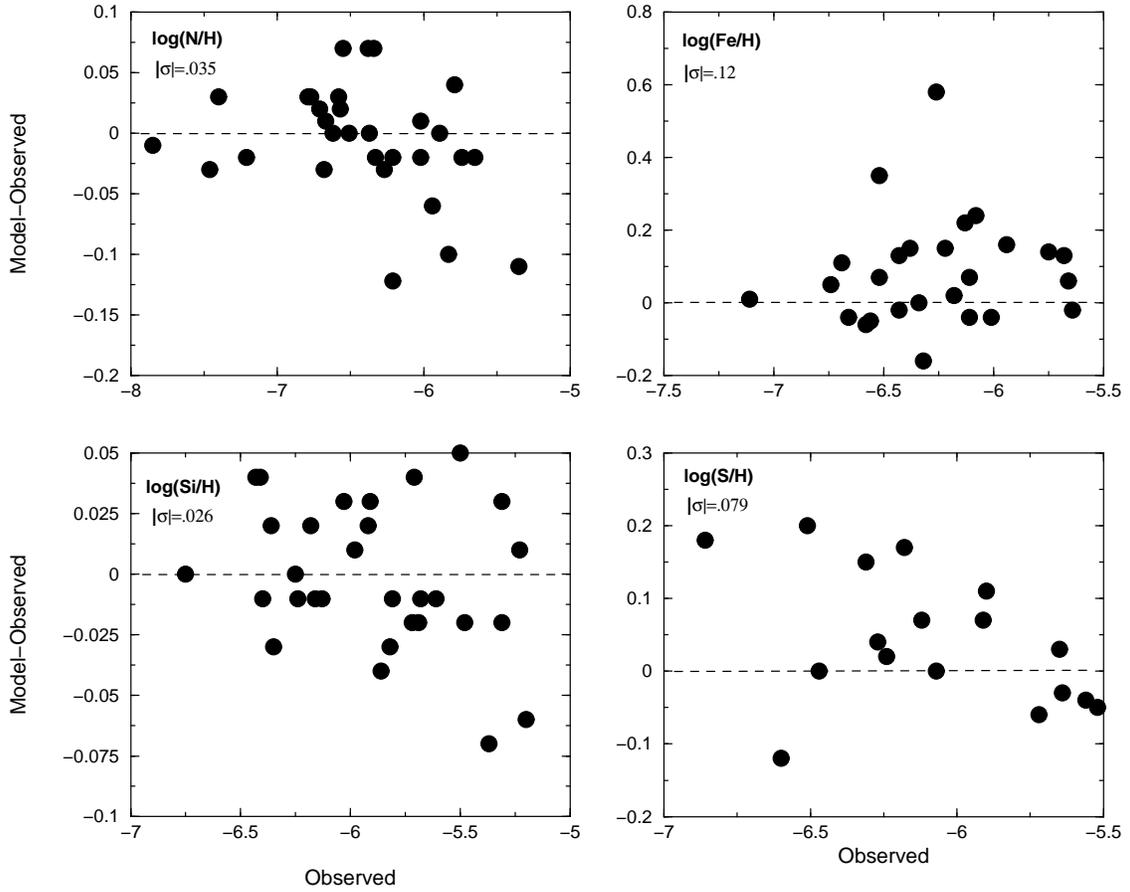}
\caption{Residuals between the model predictions and the observed values versus observed values for log(N/H), log(Si/H), log(Fe/H), and log(S/H). The absolute value of the average residual is indicated in each panel. The horizontal dashed lines show the zero residual levels.}
\label{obs_mod_compare}
\end{figure}

Fig.~\ref{obs_mod_compare} shows plots of residuals between model predictions and observed abundances versus the corresponding observed values (i.e.\ no dust corrections)
for N/H, Si/H, Fe/H, and S/H. The mean of the absolute values of the residuals is also indicated in each panel. Dashed lines indicate levels of zero residual. Again, we see that agreement between observations and the model predictions is very good in particular for N/H, Si/H, and S/H, where the average residual is less than 0.1~dex in each case. The average residual for Fe/H is only slightly larger. We also see a preference for model predictions of S/H to exceed observations, with an even larger trend in this direction in the case of Fe, which may imply mild depletion of Fe onto dust.

Finally, the uniqueness of models in Table~\ref{singlemodels} was assessed using Monte Carlo simulation techniques in conjunction with the parameter ranges inferred from our models. The test is described in detail in Appendix~B, where we conclude that, within the confines of our assumptions, our models have a high probability of being unique.

\section{DISCUSSION}

\subsection{Star Formation Efficiency and Evolutionary Age}

\begin{figure}
\centering
\includegraphics[width=12cm,angle=270]{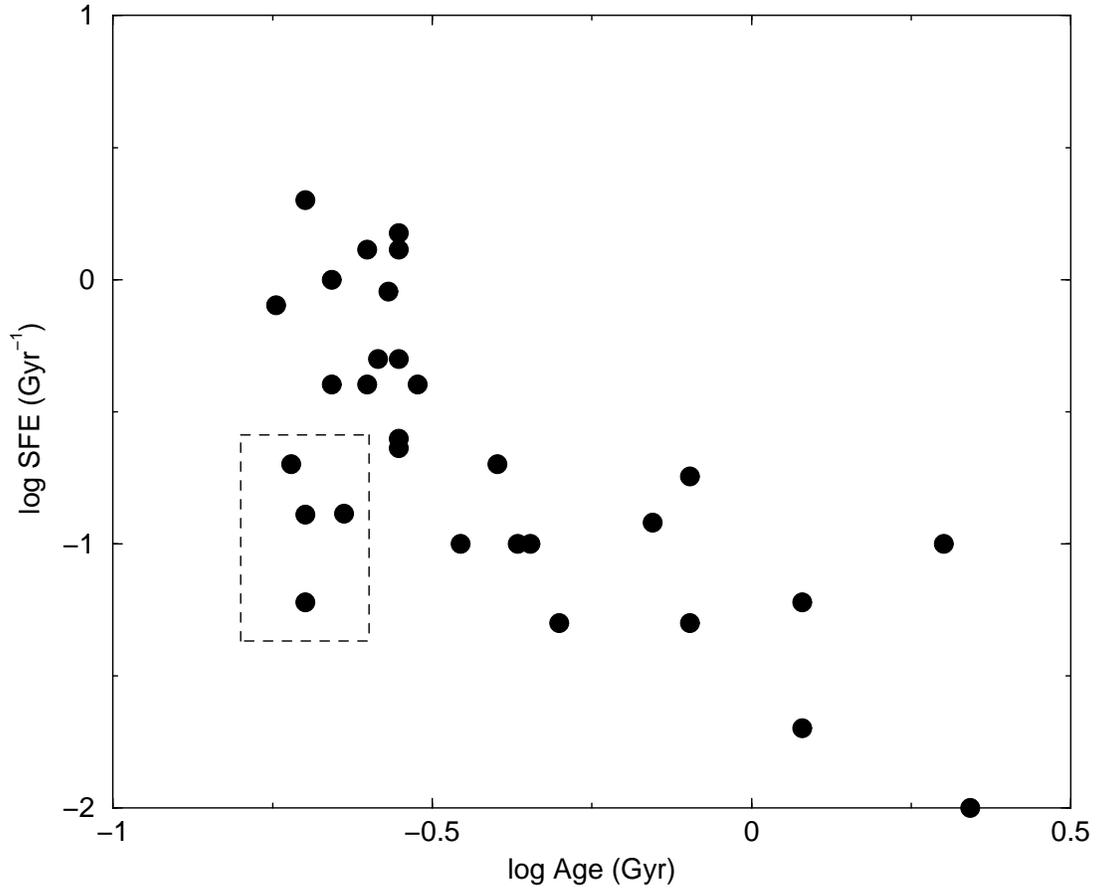}
\caption{Model-determined values for the star formation efficiency in Gyr$^{-1}$ and evolutionary age in Gyr for our sample DLAs. A box has been drawn around the low nitrogen DLAs.}
\label{sfe_age}
\end{figure}

The values for log(SFE) in Gyr$^{-1}$ and log(age) in Gyr for each object, as inferred from our models, are plotted against each other in Fig.~\ref{sfe_age}. The four LNDLAs are enclosed in a box. Note that the LNDLAs are grouped together in the region of low SFE and age, consistent with their positions in Fig.~\ref{n2sivsi} at low values of Ni/Si and Si/H.
Values for SFE range between 0.01-1.5~Gyr$^{-1}$, with a mean and median of 0.44 and 0.20 Gyr$^{-1}$, respectively. At the same time, the evolutionary ages range between 0.18-2.2~Gyr. The SFE values for DLAs are similar to those inferred for BCGs (Henry, Nava, \& Prochaska 2006), although BCGs tend to be more metal-rich (see our discussion of Fig.~\ref{n2sivsi} above). They are also in good agreement with the model results of Lanfranchi \& Matteucci (2003) and Chiappini, Matteucci, \& Meynet (2003) for BCGs and DLAs, suggesting that BCGs are more advanced than DLAs in the evolutionary sense. Clearly the DLAs in our sample cannot have evolved for a time greater than the ages implied by the redshifts of the absorbing systems. At the same time, evidence is building that indicates that BCGs contain AGB stars and are at least several Gyr in age. (See Aloisi, Tosi, \& Greggio 1999 and {\"O}stlin 2000 for the specific case of I Zw 18).

One of the most striking attributes of Fig.~\ref{sfe_age} is the apparent anticorrelation between SFE and age, which is especially strong among the objects not enclosed in the box, i.e. those which are not LNDLAs. The SFE inferred through our models is actually the {\it time-averaged} value over the age of a system, since the SFE remains constant throughout the calculation. But star formation could actually have been much more prodigious early on in many objects. This would be the case if bursts dominated at an early age, and then gave way to a more tranquil period of quiescent, continuous star formation. Supposing then that all DLAs in our sample evolved in this way, a younger object would have a higher average SFE than an older one. Ultimately, this finding could be related to the decline in surface density of star-forming materials with time as a system forms stars and evolves. {\it We conclude that the SFE declines as a DLA evolves.}

At the same time, the fact that we derive a supersolar Si/Fe value for all of our objects suggests that star formation has not ceased altogether. If it had, Si production by rapidly evolving massive stars would have stopped while Fe production from slower evolving Type~Ia supernova progenitor systems would have continued, resulting in more solar-like values for Si/Fe (e.g.\ Dessauges-Zavadsky et al.\ 2004).

Finally, the data in Fig.~\ref{sfe_age} imply that there are no objects characterized by both relatively high SFE and advanced age, suggesting that DLAs with those characteristics are either very rare or difficult to detect.

In relatively unevolved objects such as DLAs where the gas fraction is nearly unity (see eq.~A10), the star formation rate is equal to the product of the SFE and the surface mass density, i.e. $\psi \approx \nu M$. While our models help us constrain the SFE, they do not limit the surface mass density, since our predicted element-to-element ratios are not a sensitive function of that quantity. Thus, we are unable to use our results to infer unique star formation rates to compare with observed ones. However, by combining our derived SFE values with the average DLA star formation rates from Wolfe, Prochaska, \& Gawiser (2003) we can infer something about the surface densities of the DLAs within our sample.

For an average star formation rate Wolfe et al. found log$\psi$=-2.2 when a cold neutral medium (CNM) is assumed, and log$\psi$=-1.3 for a warm dark medium (WNM), where the units of $\psi$ are M$_{\odot}$/kpc$^2$/yr. In the case of a CNM then, we find a range of surface mass densities in M$_{\odot}$/pc$^2$ of 3.2 to 315.5 with average and median values of 69.1 and 31.6, respectively. Likewise, for the WNM scenario, we find the range to be 25.05-2505 with an average and median of 548.6 and 250.5. While the WNM results appear to give unrealistic values for surface mass density, the CNM results are for the most part consistent with typical values of dwarf irregulars (the low end of the density range) and spiral disks.

\begin{figure}
\centering
\includegraphics[width=12cm,angle=270]{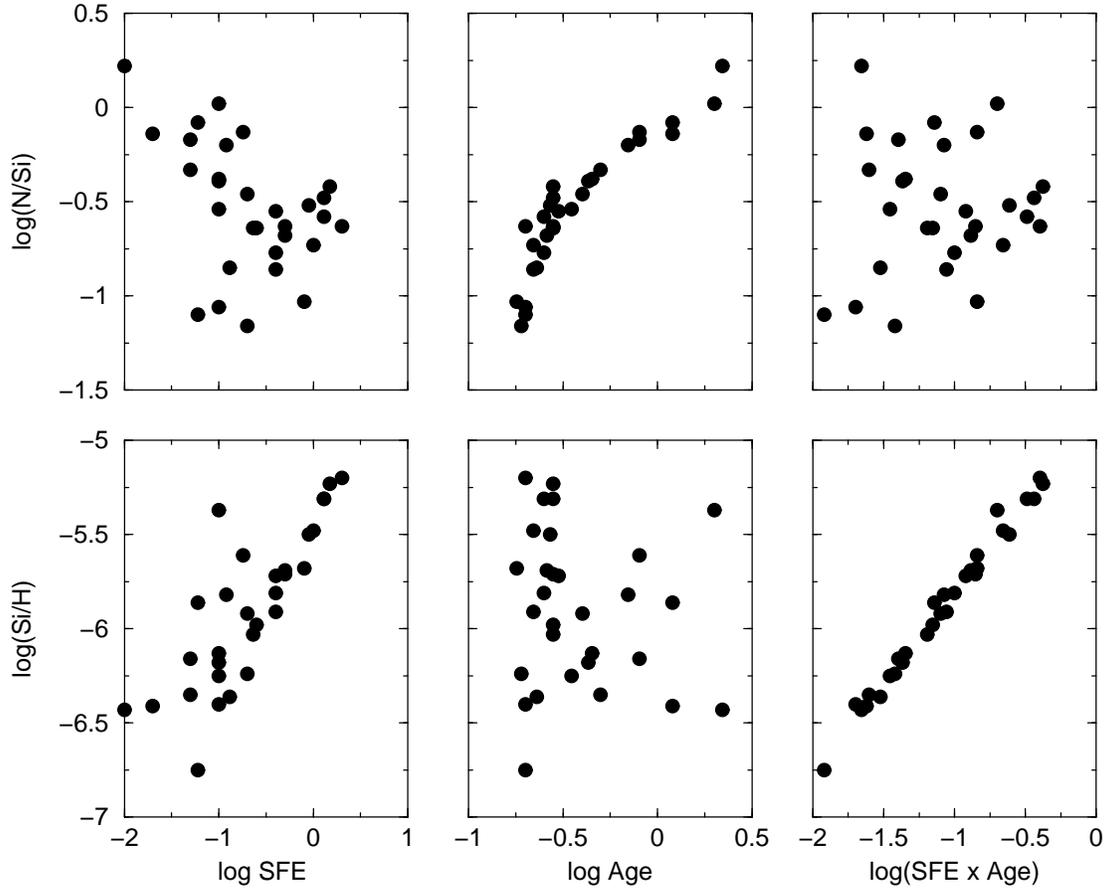}
\caption{Upper panels: Observed values of log(N/Si) versus SFE, age, and the product of these two parameters, respectively left to right. Lower panels: Same as the upper panels but for log(Si/H).}
\label{6plot}
\end{figure}

Correlations between abundance ratios on the one hand and SFE and evolutionary age on the other are important for understanding the underlying nature of DLAs. In Fig.~\ref{6plot} we plot {\it observed} values of log(N/Si) and log(Si/H), in the upper and lower panels, respectively, versus {\it model-determined} values of log(SFE), log(age), and log(SFE~x~age) left to right, respectively, as independent variables. Note that log(SFE~x~age) is related to total astration. In the upper panels, there appears to be a weak anticorrelation between log(N/Si) and log(SFE), a very strong direct one between log(N/Si) and age, and no apparent relation between log(N/Si) and log(SFE~x~age). Note also that the strong link between N/Si and age that we find is consistent with the suggestion by Edmunds \& Pagel (1978) regarding extragalactic H~II regions that N/O tracks age and is due to the sensitivity of N evolution to the delayed release of this element by LIMS as they evolve at a slower pace relative to massive stars. The weak anticorrelation between N/Si and SFE is linked to the anticorrelation described above between SFE and age. The lower three panels show a weak direct link between Si/H and SFE but a very strong one between Si/H and SFE~x~age. Since the latter quantity is related to total astration (the total number of stars formed over the lifetime of the system) and manifested in a system's metallicity, it is not surprising that SFExage is closely linked to a metallicity tracer such as Si/H. 

In computing our models, then, we relied upon the tight correlation between log(N/Si) and age to constrain the latter parameter. Once the age was determined, we could then adjust the SFE to bring the predicted value of log(Si/H) into line with its observed counterpart. Thus, despite the apparent degeneracy between SFE and age demonstrated in Fig.~\ref{sfe_age}, we were able to infer a unique set of input parameters for each object in this way.

Only a small number of chemical evolution models of DLAs have been published, so a comparison with others' results is very limited. Dessauges-Zavadsky et al. (2004) modeled three objects, including Q1331-17, 
which is also present in our sample. These authors calculated both a 
spiral galaxy and a dwarf galaxy model, with the latter being more directly 
comparable to ours for this object. They derived a value for 
SFE of 0.03~Gyr$^{-1}$, compared with our finding of 0.06, along with an age
$\gtrsim 3.5$\,Gyr compared to our value of 1.2~Gyr. Their greater age was necessary to match the near-solar Si/Fe value which they inferred after
applying a substantial dust correction to the observed Si/Fe ratio.

In a more recent paper which appeared after our study was nearly complete, Dessauges-Zavadsky et al. (2007) used the same approach as in their 2004 paper to model six more objects, four of which are present in our list, i.e. Q0841+129 (z=2.375), Q1210+17, Q2230+02, and Q2348-1444. Comparing our SFE and age results with the same parameters from their favored models the two studies are consistent for all but the case of Q2230+02, where we derive a much larger SFE and much smaller age.

Finally, whatever the details are regarding the physical systems hosting DLAs, their chemical evolution is currently comprehensible simply through the use of very basic one-zone models characterized by low continuous star formation rates and only two variable parameters, namely the system's SFE and evolutionary age. At this point in our understanding of these objects, no other specifications appear to be needed. 
{\it DLA abundances apparently can be reduced to a two dimensional problem, as illustrated in Fig.~\ref{6plot}.}

\subsection{Metal Abundance Ratios}

Metal abundance ratios such as N/Si, Si/Fe, and S/Si are influenced by the effective yields of the two metals making up the ratio as well as the relative rates of evolution of the stellar groups responsible for their production. Such ratios may also be sensitive to dust depletion if one of the elements is more refractory than the other. Problems associated with N/Si and Si/Fe were summarized briefly in the Introduction, where we showed that questions surrounding their behavior in DLAs, such as the N/Si scatter and possible bimodality as well as the ubiquitous supersolar Si/Fe values, are as yet unresolved.
Here we employ the results of our models to address some of the questions associated with these abundance ratios. 

\subsubsection{N/Si Range and Bimodality}

The large scatter in N/Si described in \S\ 2 is most likely the result of age differences among our sample objects.
Given the tight link between N/Si and age exhibited by our models, we conclude that the scatter in N/Si is evidence of an age range among DLAs, due to the delay in N release by LIMS resulting from their slower evolution. Thus, the LNDLAs may actually be objects observed at a very young evolutionary age, i.e. a small fraction of a Gyr. Initially in these objects the N/Si climbs because of massive star N yields, but as time progresses, N produced by LIMS is released during their final evolutionary stages, and N/Si rises to greater levels.

\begin{figure}
\centering
\includegraphics[width=12cm,angle=270]{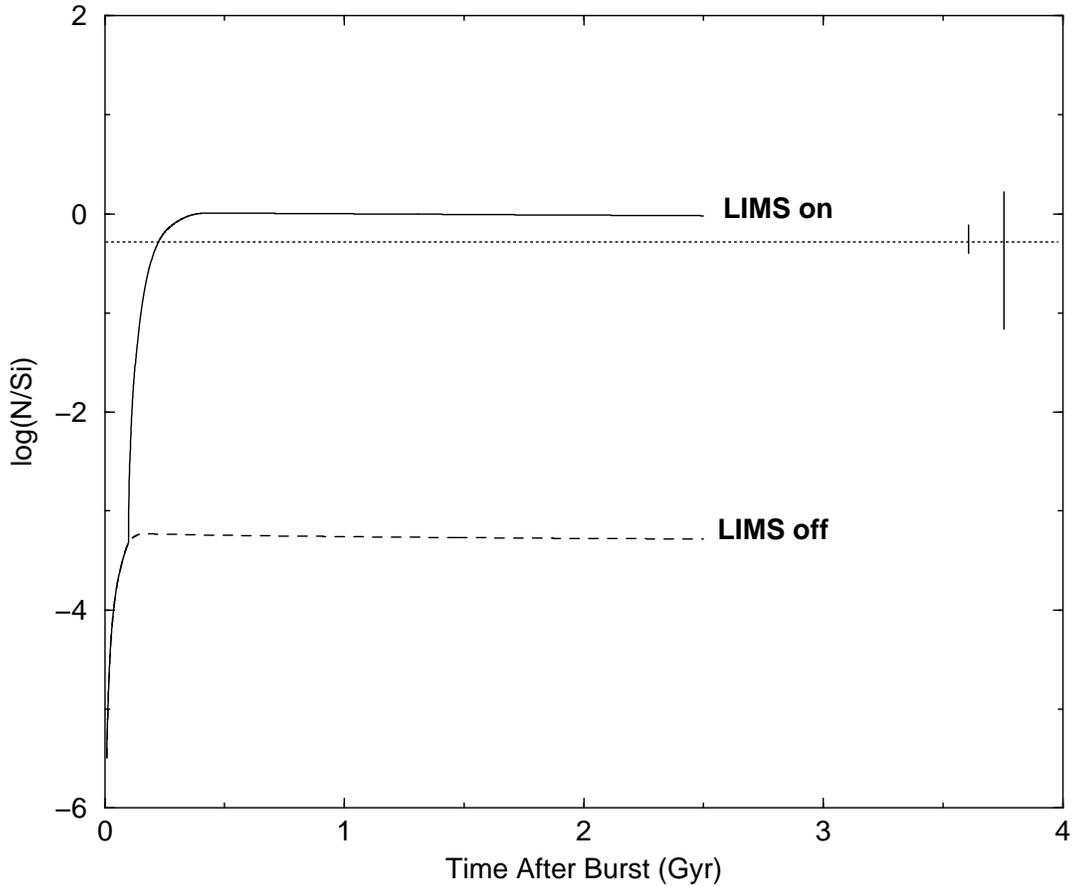}
\caption{log(N/Si) versus the time in Gyr following a starburst lasting 100~Myr. The solid line relates to a model in which both massive stars and LIMS contribute to N production, while the short dashed line shows the result for a similar model except that N production by LIMS is turned off. The long horizontal dotted line indicates the average log(N/Si) for plateau blue compact galaxies, using log(N/O) from Nava et al. (2006) and converting to log(N/Si) using the solar values in Asplund et al. (2005). The shorter of the two vertical lines on the right shows the BCG range, while the longer one shows the range for DLAs.}
\label{n2si_vs_time}
\end{figure}

This idea is illustrated in Fig.~\ref{n2si_vs_time}, where we show the time evolution of N following a burst of 100~Myr duration. The solid line represents the model with both massive stars and LIMS contributing to N buildup, while the short-dashed line shows the track followed by a system in which the N production by LIMS has been turned off. The dotted line shows the average value for blue compact galaxies on the plateau at -0.28, as inferred by Nava et al. (2006) for N/O and converted to N/Si here using the solar O/Si ratio from Asplund et al. (2005). The actual range in BCG values extends from -0.39 to -0.12 and is indicated with the short vertical line.  Recall that the observed range for DLAs of log(N/Si) is -1.16 to +0.22, shown in the figure by the longer vertical line. For both models, we employed the same code as the one described above with SFE=0.01 Gyr$^{-1}$. The solid curve turns over at roughly the location of the BCG value at a time of 230~Myr and levels off at 400~Myr. HEK estimated from their models that most N release has occurred by 250~Myr, although they used the LIMS yields of van den Hoek \& Groenewegen (1997) which produce more N than Marigo's (2001) predictions (see comparison of effective yields of LIMS models by Henry 2004). In any case it is clear that LIMS contribute significantly to N buildup, according to our models and our selection of stellar yields. 

We also conclude from Fig.~\ref{n2si_vs_time} that BCGs have generally been evolving longer than DLAs, since their vertical range is much smaller, and the mean N/Si for the objects on the plateau is close to the value at which N/Si of the bursting model above levels off. As already pointed out in \S\ 4.1, this conclusion is consistent with other evidence indicating that BCGs are at least several Gyr in age.

Data for N/Si and Si/H presented in Prochaska et al. (2002) was the first to indicate that the distribution of DLAs in the N/Si--Si/H plane may be bimodal. We saw evidence of this in Fig.~\ref{n2sivsi}, where a diagonal gap extending from upper right to lower left appears to separate the four LNDLAs inside the box from the main body of objects.

\begin{figure}
\centering
\includegraphics[width=12cm,angle=270]{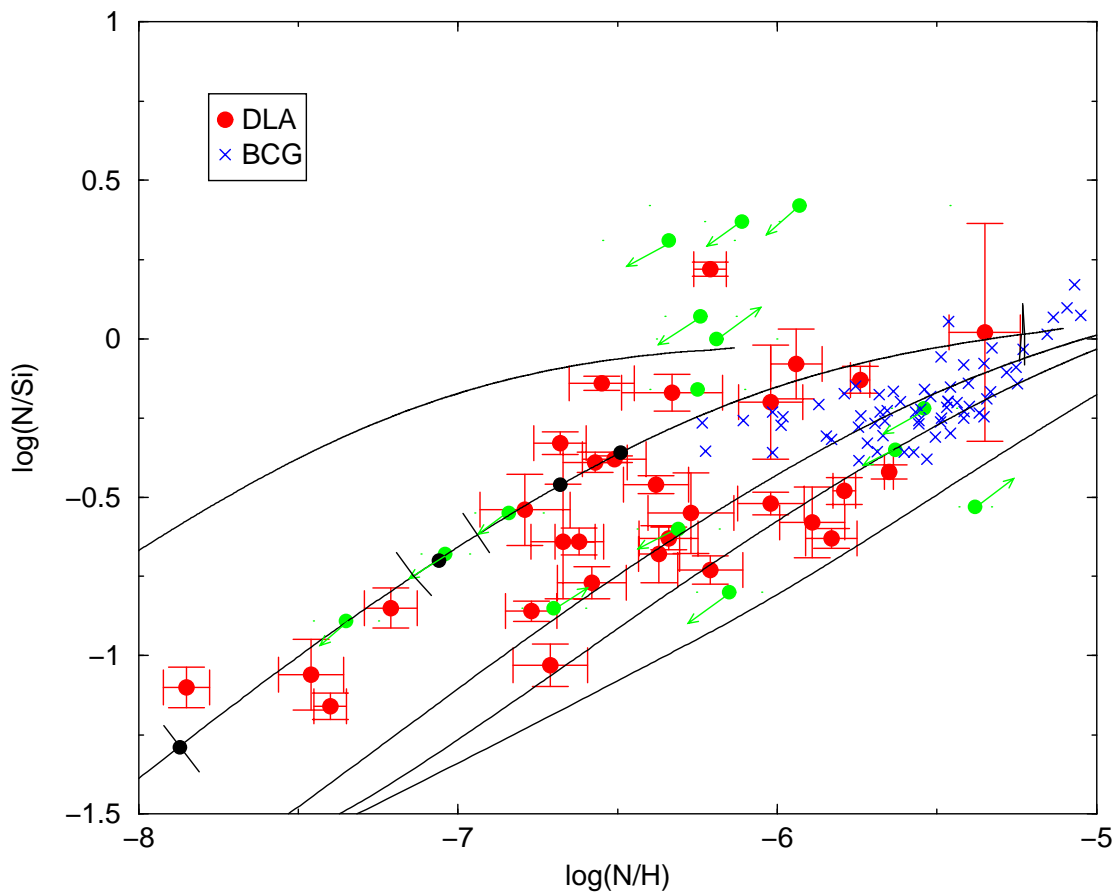}
\caption{Observed values of log(N/Si) versus log(N/H) for our sample of DLAs (red circles) and the blue compact galaxies (blue crosses) from Nava et al. (2006). DLA values where only limits are available are shown with green dots with arrows. The five model tracks show predicted evolution through the plane for different star formation efficiencies, where the values of the latter in Gyr$^{-1}$, upper left to lower right, are 0.01, 0.10, 0.50, 1.0. and 4.0. Time steps in increments of 0.1~Gyr are shown with black circles on the track corresponding to SFE=0.10. This same track is divided into three segments by the short bars perpendicular to the track (see text).}
\label{n2sivn2h}
\end{figure}

This gap is seen even more clearly in Fig.~\ref{n2sivn2h}, where we now plot log(N/Si) versus log(N/H). Here the DLA positions are shown with red circles along with the data for blue compact galaxies from Nava et al. (2006) indicated with blue crosses. DLA measurements which are only limits are shown with green circles and arrows. The four LNDLAs can be seen in the lower left, separated from the main group of DLAs by an obvious gap at log(N/H)~$\approx -7$. 

There are two basic hypotheses which would explain such a bimodality\footnote{We respectfully acknowledge that some of our colleagues believe that the gap in Fig.~\ref{n2sivn2h} is the result of small number statistics, that the gap will disappear once more data are available, and therefore the analysis which follows is moot. While we certainly agree that this is a possibility, we also feel that until that day comes, we need to explore the potential explanations for the gap in the event that more data show it to be real. If future observations prove that the gap-skeptics are correct, then we'll happily cease our wild speculations.}. 
First, it may simply be a statistical artifact which occurs when a small number of objects in the N/Si--Si/H plane are selected from a large parent population whose members are distributed continuously but with non-uniform density within that plane. Or second, the LNDLAs could represent a group of objects which evolved with a different set of effective yields than the rest of the objects. 

The first hypothesis is related to the explanation originally put forth by Molaro (2003) and Molaro et al. (2003). That is, the LNDLAs are objects whose N/Si evolution has temporarily been suspended until their LIMS begin releasing N. Once that begins, these objects move upward rapidly. Furthermore, with a small sub-population such as ours, a gap separating the LNDLAs and the main body of objects might be likely to appear, since the probability of finding an object there is small due to rapid evolution through that area.

We attempted to evaluate this hypothesis by assuming that the parent population of DLAs, of which our objects are a subsample, are smoothly distributed in age, i.e. there were no times in the past when the formation of systems harboring DLAs was more favored than at other times. We calculated five model tracks which differed only in their input SFE. These tracks span the entire observed set of objects and are shown in Fig.~\ref{n2sivn2h}. The SFE values  were (upper left to lower right) 0.01, 0.10, 0.50, 1.0, and 4.0 Gyr$^{-1}$. Since the track corresponding to SFE=0.10 runs through the LNDLAs, we chose it for use in our analysis. Ignoring all objects with upper/lower limits, all DLAs lying close to this track share roughly the same SFE but have different ages, 
where age increases along the track from lower left to upper right. 
We have placed small black circles at intervals of 0.1~Gyr to show how progress changes with age. 
If the DLA population has a uniform distribution of ages,
then one would expect the same number of objects per interval of spacing of the
black circles.

We have also divided the track into three segments indicated by the four short perpendicular bars placed along it. The lengths of the segments, beginning in the LNDLA region, correspond to 85~Myr, 42~Myr, and 3~Gyr, respectively. There are four objects near the track in the first segment, none in the second, and eight in the third. Based on the number of objects along the first segment we would expect to find two objects in the gap, where none actually occurs. On the other hand, a lack of objects in the gap is very consistent with the observation of only eight objects along the 3~Gyr segment. Note that along the 0.10 track near the gap there is one object with a N upper limit in the gap and one more like it in the direction of greater N/Si. The first point could very well be a LNDLA, while the second one could potentially occupy the gap. However, even if this were the case, the results of our test would be unchanged. Thus, our test fails to rule out the possibility that the gap is simply a statistical artifact. 

We now consider the second hypothesis. 
In their discussion of the N/Si bimodality, Prochaska et al. (2002) used chemical evolution models to demonstrate that this phenomenon could be explained by a non-standard IMF. They proposed that the LNDLAs actually represent a separate population in which star formation is characterized by either a highly flattened IMF or an IMF which is truncated below a certain mass threshold. Both cases result in a lower proportion of LIMS in the stellar population, and hence the integrated effective N yield is reduced.

In order to check on the feasibility of this second hypothesis, we calculated additional individual models for the four LNDLAs in our sample, i.e. Q0151+0448, Q1008+36, Q2206-19, and Q2348-14. These models were generally identical to the ones presented in Table~\ref{singlemodels} except that their IMF properties were altered. In the first instance the IMF was truncated below 4.80\sun (or 4.65\sun for Q2348-14), while in the second instance the IMF slope was assumed to be 0.10 (recall that the IMF slope for the Salpeter relation is 1.35). We then proceeded as before in the general modeling, using log(N/Si) and log(Si/H) as our primary target values.

Results of this experiment are presented in Table~\ref{altimf}, where the format of Table~\ref{singlemodels} is adopted. Note that the results for the LNDLAs from Table~\ref{singlemodels}, using a normal IMF, have been duplicated here for ease of comparison.

For the case of a truncated IMF, successful matches to N/Si and Si/H were obtained for all four objects. While the IMF for three of the models was truncated below 4.8\sun, in the case of Q2348-14, whose observed value of N/Si is somewhat higher than the same ratio for the other three, the threshold had to be lowered slightly to 4.65\sun\ to provide more N to the integrated yield. For the flat IMF, N/Si and Si/H were well-matched for three of the LNDLAs but a good model of Q1008+36 proved unattainable. Of the four LNDLAs, this object possesses the lowest N/Si value but the highest level of Si/H. Despite numerous trials with a wide variety of SFE and age input values we could not find a suitable combination.

In comparing the SFEs and ages for the three IMF forms represented in Table~\ref{altimf} we see that truncated IMF models in general require longer evolutionary ages than either the flat or normal IMF examples, since additional time is required to produce the N that would have been otherwise synthesized by LIMS. Another trend is that both the truncated and flat IMF cases tend to have smaller SFE values.

We note that Lanfranchi \& Matteucci (2003) found that when an altered IMF is used to reduce the N/Si to the level of the LNDLAs, Si/Fe becomes unacceptably high. This point will be discussed in more detail in the next subsection, where we show that Si/Fe in our models does indeed change when the IMF is altered, but not by untenable amounts.

Therefore, the flattened or truncated IMF hypothesis appears to be capable of explaining the LNDLAs.
Many theoretical studies of Population~III stars clearly predict that the first stellar generation either had a greater proportion of massive stars or its IMF was truncated in such a way that the relative number of LIMS was small (see the review by Bromm \& Larson 2004). Objects possessing such a stellar population are likely located at high redshifts and formed when metallicities were below 10$^{-4}$Z$_{solar}$ (Schneider et al. 2002). Komiya et al. (2007) conclude that the changeover from a top-heavy to a normal IMF occurred around [Fe/H] of roughly -2.5, i.e. within a dex of the metallicities found for the LNDLAs. Our analysis thus suggests that some DLAs may exhibit the chemical imprint expected from Population~III stars.

\subsubsection{The Supersolar Si/Fe Ratio}

\begin{figure}
\centering
\includegraphics[width=12cm,angle=270]{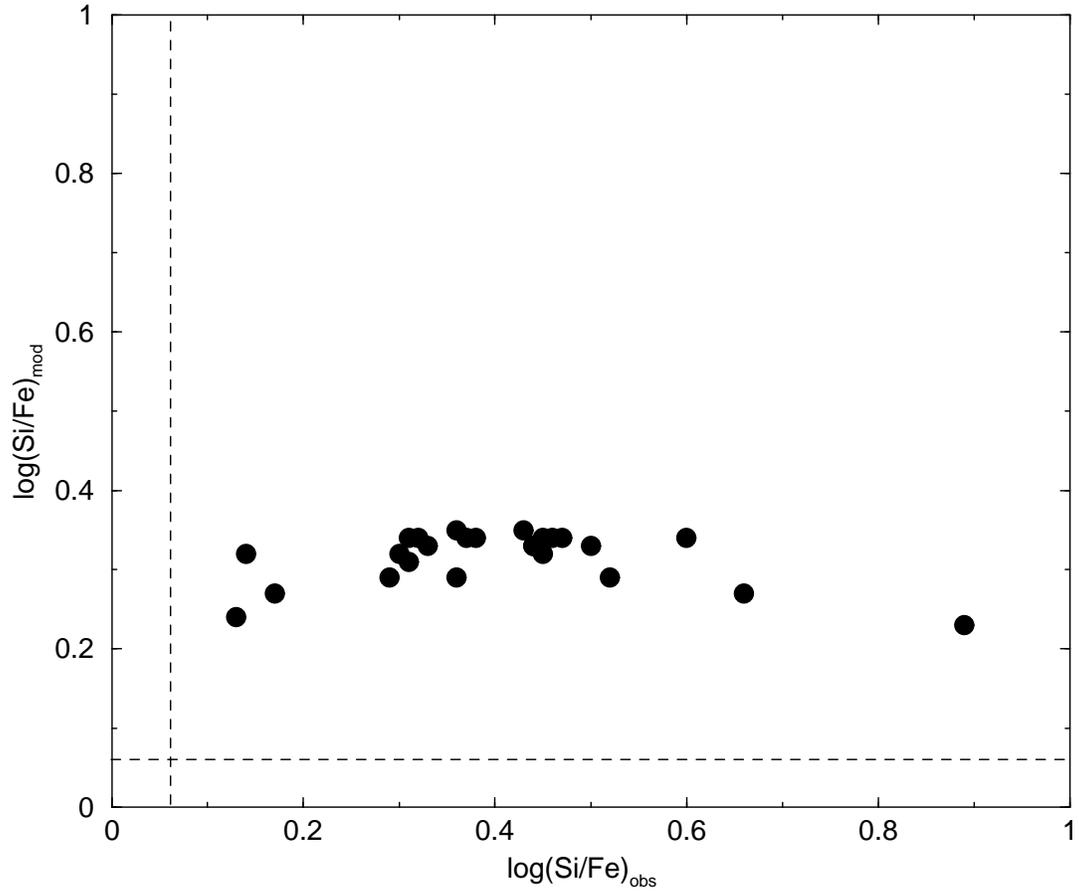}
\caption{log(Si/Fe) values from models (vertical axis) versus observed values (horizontal axis). The dashed lines show solar values in each direction, from Asplund et al. (2005).}
\label{si2fe}
\end{figure}

Another important chemical diagnostic of DLAs is the Si/Fe ratio. Since Fe is more refractory than Si (Savage \& Sembach 1996), the observed gas phase Si/Fe ratio may tend to be higher than the intrinsic value. 
Indeed, DLAs exhibit significantly larger Si/Fe ratios at higher
Si/H (Figure~2), which is explained by differential 
depletion (Prochaska \& Wolfe 2002).
For the majority of DLAs in this paper, however, the metallicity is low
and the depletion effects may be small.
In \S\ 2, we pointed out the tendency for observed Fe abundances to be 
$\approx 0.1$\,dex less than their 
predicted counterparts in Fig.~\ref{obs_mod_compare}, upper right panel. 
This is suggestive of a mild, but non-negligible depletion effect. Note that all of our models predict supersolar values for Si/Fe. The average value of Si/Fe predicted by our models is +0.31 ($\sigma$=.034) for the entire sample, the ratio being roughly constant overall. Fig.~\ref{si2fe} is a plot of Si/Fe for the models versus the observations. The dashed lines show the solar value (Asplund et al. 2005) for each axis. Notice that the observed values are systematically higher than the predicted ones and show a much broader range. 

Our models indicate that the supersolar Si/Fe ratios measured in most DLAs stem from their relatively unadvanced ages. Roughly half of all Fe is produced in massive stars, while the other half comes from Type~Ia supernovae, the precursors of which evolve more slowly than massive stars. As a result, Fe release is partially delayed over a timescale of between 40-50~Myr for bursting systems or 4-5~Gyr for spiral disks (Matteucci \& Recchi 2001). Therefore, the supersolar Si/Fe values observed in DLAs is expected because of their generally young evolutionary ages. On the other hand, the cause of the {\it scatter} in Si/Fe exhibited by DLAs is more difficult to assess, since both an age range and depletion of Fe onto dust are likely to play roles in determining Si/Fe.

In our discussion of the N/Si ratio above, we pointed out that
one objection to the IMF modification hypothesis was raised by Lanfranchi \& Matteucci (2003), who predicted log(Si/Fe) values of roughly +0.6, nearly 0.25~dex above observed levels, for a model with an IMF slope of 0.10 while at the same time predicting log(N/Si) levels which were consistent with those of the LNDLAs. Qualitatively, an increase in Si/Fe is expected, because a flattening of the IMF slope reduces the relative contribution of Type~Ia supernovae to the effective Fe yield, since these explosive events originate from stars of generally lower mass than the stars which produce Si. As a result the Si/Fe ratio should be higher in the case of a flattened IMF.


We investigated this point in some detail, as it may represent a way to eliminate the altered IMF hypothesis, thereby clarifying the cause of the N/Si bimodality to some extent. We ran models using three different IMF forms, i.e. Salpeter ($\alpha$=1.35), flattened ($\alpha$=0.10), and truncated (M$_{trunc}$=4.8\sun with $\alpha$=1.35). For each IMF form we used three different values for the SFE, 0.003, 0.05, and 1.0 Gyr$^{-1}$. In all cases the calculated value of Si/Fe was consistent with the observations, and always remained well below the values predicted by Lanfranchi \& Matteucci. We speculate that the different results are related either to the different yield sets employed or to differences in star formation prescriptions, e.g. bursting rather than continuous star formation. We note, however, that regardless of the IMF form that we used, we found that values did not differ from observations by more than about 0.10~dex and thus within the observational uncertainty of log(Si/Fe). Therefore, we currently see no reason to dismiss either the flattened or truncated IMF hypotheses for explaining LNDLAs because of a failure to predict observed Si/Fe values. 


In summary, our models employing continuous star formation predict supersolar values for Si/Fe that fall within a range which is much narrower than the broad one seen observationally. The most likely explanation is Fe depletion onto dust, although bursting star formation histories, where burst length and frequency both vary from object to object, may also
be capable of explaining the observed scatter in Si/Fe.

\subsubsection{S/Si}

The filled circles in Fig.~\ref{svsi} illustrate the tight relation between Si and S observed in DLAs. Model results from Table~\ref{singlemodels} are plotted in the same figure as open circles, where it is seen that they all fall very close to the least squares fit (straight line) to the data. Thus, observation and theory closely match up in this case.

The lockstep behavior between Si and S was reported by Prochaska \& Wolfe (2002) and is consistent with the generally accepted belief that these two elements are forged within the same types of stars. In addition, it indicates that Si is not significantly depleted by dust formation, at least within the metallicity range of our object sample. The results of Savage \& Sembach (1996) suggest that while Si can form grains, S generally is not found to be depleted in the interstellar medium. If Si depletion were present but unaccounted for in the DLA data, then we would expect to find a leftward horizontal offset of the data with respect to the models. Clearly, this is not seen. This absence of Si depletion in our data would appear to be in conflict with the findings of Dessauges-Zavadsky et al. (2006), who detected a positive correlation between [Si/S] and [Zn/Fe], where the second ratio is purported to increase directly with dust depletion.  
The difference may have to do with the fact that we consider the
total abundances for a set of DLAs whereas Dessauges-Zavadsky et al. (2006) examined
individual `clouds' within a DLA.   The latter analysis will be
more sensitive to the depletion effects if there are even
modest variations in Si/S and Zn/Fe within a given DLA.
Furthermore, our sample includes only a few DLAs with [Si/H]~$> -1$ where
dust effects become more pronounced (Prochaska \& Wolfe 2002).


\section{CONCLUSIONS}

We have compiled published abundances of N, Si, S, and Fe for 45 damped Lyman alpha systems in order to study trends present in these cosmologically important objects. Our attention was focused particularly on understanding the reasons for: (1)~the wide range in values of N/Si, their generally low level compared with blue compact galaxies, and the apparent bimodality in their distribution; (2)~the low metallicity of DLAs as traced by Si; and (3)~the ubiquitous supersolar Si/Fe values observed. Our primary method of analysis was to calculate an individual numerical chemical evolution model for each DLA while attempting to keep the number of assumptions about the nature of these objects as well the variable input parameters at a minimum. N and Si abundances were our principal targets for modelling, and thus we focused our attention on those 30 DLAs for which firm abundance values, as opposed to limits, were available. We have reached the following seven conclusions.  

\begin{enumerate}

\item The abundances of N, Si, S, and Fe in DLAs are easily explained by a one-zone model scheme with infall but no wind, continuous star formation, and only two variable parameters, i.e. star formation efficiency and the period of time over which evolution occurs (evolutionary age). For now, this 2-dimensional scheme appears to be sufficient, and there seems to be no need to increase the level of model complexity.

\item The ranges in SFE and age for the 30 sample objects modeled are, respectively, 0.01-1.5~Gyr$^{-1}$ and 0.18-2.0~Gyr. The mean of the SFE values is 0.44 Gyr$^{-1}$, while the median value for this parameter is 0.20 Gyr$^{-1}$. These values agree well with those found in previous model studies of dwarf irregulars as well as spiral disk locations at large galactocentric distance. 

\item There is a clear anticorrelation between SFE and age for DLAs. Since our models infer only the {\it time-averaged} SFE, we suggest that star formation in many DLAs may have occurred via bursts early on, but the intensity and/or frequency of these episodes has subsided with time. In this way older objects will tend to display lower average SFEs. Alternatively, the inverse behavior between these two parameters could be due to a steady reduction in the interstellar materials necessary for forming stars. 

\item The broad range in N/Si for DLAs compared with blue compact galaxies is explained primarily by the fact that we are observing the former at various stages of N enrichment. This difference between the two object types implies that DLAs are significantly younger in evolutionary age than are the blue compact galaxies.

\item The cause of the apparent bimodality in the N/Si distribution of DLAs is ambiguous, as it can be explained either as a statistical result or by the presence of two different sets of effective yields, one associated with a normal IMF and another associated with a truncated or flat IMF. Such top-heavy IMFs
are qualitatively consistent with current theoretical predictions about the mass distributions among Pop~III stars, suggesting that abundances in the LNDLAs exhibit the imprint of Pop III nucleosynthesis.

\item The Si/Fe ratio in DLAs exceeds the solar value for all objects in our sample. Thus DLAs exhibit a characteristic alpha enrichment also seen in globular clusters and halo stars within the Milky Way. However, our models are unsuccessful in explaining the large observational scatter in Si/Fe, and we suggest that this may be due to Fe depletion onto dust in varying amounts or bursting star formation histories in which both burst length and frequency vary among DLAs.

\item Si and S in DLAs increase together in lockstep, an observational result that is supported by our models. Since silicon's abundance is more likely to be depleted through grain formation, the observed lockstep behavior strongly suggests that in most DLAs with metallicities similar to those in our sample, Si depletion is not important. 

\end{enumerate}

The question of a N/Si bimodality can only be resolved by increasing the size of the sample of DLAs with N and Si measurements. More objects will allow us to assess the distribution in more detail and hopefully establish whether an apparent bimodality is simply the result of a random distribution of a small number of objects or is truly related to differences in effective nitrogen yield. Finally, we need to continue to explore the connection between DLAs and nearby galaxies. One possible way of searching for such connections is to begin with successful chemical evolution models of DLAs and evolve them further in time, making predictions both about elemental abundances as well as the integrated spectral properties of the more mature galaxy. A comparison of this second generation of models with current observations of suspected DLA endpoints such as spiral galaxies and dwarf irregulars might reveal an evolutionary connection.

\acknowledgments

RBCH is grateful to UCSC and Lick Observatory for the nice hospitality provided during a visit there while this project was being formulated. Likewise, JXP thanks the Homer L. Dodge Department of Physics \& Astronomy at the University of Oklahoma for an equally generous welcome during his visit there. We also thank Max Pettini for offering comments on this paper, Mercedes Moll{\'a} for helping us check our code, and Tim Young for a valuable discussion concerning Population~III stars. Both authors are grateful to the NSF for support, RBCH from grant AST03-07118 to the University of Oklahoma and JXP from grants AST 03-07824 and AST 05-48180 to the University of California, Santa Cruz. Finally, we dedicate this paper to the memory of Bernard Pagel, our friend and mentor who has recently passed away. Bernard's exacting science and encyclopedic mind inspired many of us, and it was an honor to have worked with him.

\appendix

\section{THE CHEMICAL EVOLUTION MODEL}

Our numerical code is an expanded version of the one employed by HEK. It is a one-zone chemical evolution program
which follows
the buildup of the elements H, He, C, N, O, Ne, Si, S, Cl, Ar, and Fe over time. 

We imagine a generic galactic region with a cross-sectional area of 1~pc$^{-2}$ that is initially empty but then accretes matter by way of infall of pristine, metal-free gas whose rate is given by:
\begin{equation}
f(t)=\Sigma_{t_o}\left\{\tau_{scale}\left[1-exp\left(-\frac{t_o}{\tau_{scale}}\right)\right]\right\}^{-1}exp\left(-\frac{t}{\tau_{scale}}\right) M_{\odot}Gyr^{-1}pc^{-2},
\end{equation}
where $t_o$ and $\Sigma_{t_o}$ are the current epoch and surface density, the latter is taken to be 5~M$_{\odot}$pc$^{-2}$, $\tau_{scale}=7$~Gyr, and $t_o$=13.7~Gyr\footnote{All models assumed the same surface density of 5~M$_{\odot}$pc$^{-2}$, which is consistent with observed values for blue compact galaxies, one of the proposed host types for DLAs (L.~van~Zee, private communication).}. This form of the infall rate is the one discussed and employed by Timmes, Woosley, \& Weaver (1995) for their models of the Galactic disk.

As star formation commences, 
the total mass M within the region becomes partitioned into interstellar gas of 
mass $g$ and stars of mass $s$ such that
\begin{equation}
M=g+s.\label{mgs}
\end{equation}
If $\psi(t)$ and $e(t)$ are the rates of star formation and stellar ejection, 
respectively, then:
\begin{equation}
\dot{s}=\psi(t)-e(t)
\end{equation}
and
\begin{equation}
\dot{g}=f(t)-\psi(t)+e(t).\label{g}
\end{equation}
The interstellar mass of element $x$ in the zone is $gz_x$, whose time
derivative is:
\begin{equation}
\dot{g} z_x + g \dot{z}_x = -z_x(t) \psi(t) + z^{f}_x f(t) + e_x(t),
\end{equation}
where $z_x(t)$ and $z^{f}_x$ are the mass fractions of $x$ in the
gas and in the infalling material, respectively, and $e_x(t)$ is
the stellar ejection rate of $x$. Solving for $\dot{z}_x(t)$ yields:
\begin{equation}
\dot{z}_x(t)=\{f(t)[z^{f}_x - z_x(t)] + e_x(t) -e(t)z_x(t)\}g^{-1}.
\end{equation}
The second term on the right hand side of eq.~A6 accounts for the injection of metals into the gas by stars, while the first and third terms account for effects of dilution due to infall and ejected stellar gas, respectively.

The rates of mass ejection $e(t)$ and ejection of element $x$, $e_x(t)$, are:
\begin{equation}
e(t)=\int_{m_{\tau_{m}}}^{m_{up}}[m-w(m)]\psi(t-\tau_m)\phi(m)dm\label{e}
\end{equation}
and
\begin{equation}
\begin{array}{lr}
e_x(t)=(1-C)\int_{m_{\tau_{m}}}^{m_{up}}\{[m-w(m)]z_{x}(t-\tau_{m})+mp_x(m,z_{t-
\tau_{m}})\}\psi(t-\tau_m)\phi(m) \ dm\\
\\
+ 
C\int^{M_{BM}}_{M_{Bm}}p_x(m,z_{t-\tau_m})m\phi(m)\int^{.5}_{\mu_{min}}24\mu^2
\psi(t-\tau_{m_2})d\mu\ dm_B. \label{ex}
\end{array}
\end{equation}
In Eqs.~\ref{e} and \ref{ex}, $m$ and $z$ are stellar mass and metallicity, 
respectively, and $m_{\tau_m}$ is the turn-off mass, i.e. the stellar mass whose 
main sequence lifetime corresponds to the current age of the system. This 
quantity was determined using results from Schaller et al. (1992). $m_{up}$ is 
the upper stellar mass limit, taken to be 120~M$_{\odot}$, $w(m)$ is the remnant 
mass corresponding to ZAMS mass $m$ and taken from Yoshii, Tsujimoto, \& Nomoto 
(1996). $p_{x}(z)$ is the stellar yield, i.e. the mass fraction of a star of 
mass $m$ which is converted into element $x$ and ejected, and $\phi(m)$ is the 
initial mass function. Our choice of stellar yields is discussed below.  
In eq.~\ref{ex} the first integral gives the contributions to the ejecta of single stars, 
while ejected masses of C, O, Si, S, Cl, Ar, and Fe by SNIa through binary star 
formation are expressed by the second integral, where our formulation follows 
Matteucci \& Greggio (1986). Assuming that the upper and lower limits for binary 
mass, M$_{BM}$ and M$_{Bm}$, are 16 and 3~M$_{\odot}$, respectively, 
eq.~\ref{ex} splits the contributions from this mass range between single and 
binary stars. The relative contributions are controlled by the parameter $C$ 
which we take to be equal to 0.05 when $3\le M \le 16\ M_{\odot}$ and zero 
otherwise. The variable $\mu$ is the ratio of the secondary star mass $m2$ to 
the total binary mass, $m_B$, where $\mu_{min}=max(1.5,m_{\tau_m},m_B-8)$.

The initial mass function $\phi(m)$ is the normalized Salpeter (1955) relation
\begin{equation}
\phi(m)=\left[\frac{1-\alpha}{m_{up}^{(1-\alpha)}-m_{down}^{(1-\alpha)}}\right] 
m^{-(1+\alpha)},\label{psi}
\end{equation}
where $\alpha$=1.35 and $m_{down}=1$~M$_{\odot}$.
Finally, the star formation rate $\psi(t)$ is given by
\begin{equation}
\psi(t)=\nu M \left(\frac{g}{M}\right)^{1.5}\label{sfr}
\end{equation}
where $\nu$ is the star formation efficiency in Gyr$^{-1}$. The form of eq.~A10 is taken from Timmes, Woosley, \& Weaver (1995).

Our calculations assumed a timestep length of one million years, a value which is
less than the main sequence lifetime of a star with a mass equal to $m_{up}$, or 120~M$_{\odot}$.

At each time 
point, the increment in $z_x$ was calculated by solving eq.~A6 and 
the required subordinate equations \ref{e} and \ref{ex}. This increment was 
then added to the current value and the program advanced to the next time step. 
Finally, the total metallicity at each point was taken as the sum of the mass 
fractions of all elements besides H and He. The unit of time is the Gyr, while 
the mass unit is the solar mass.

Single star yields for all elements but Cl and Ar were taken from Portinari, 
Chiosi, \& Bressan (1998)\footnote{We note that the quantities published by Portinari 
et al. represent the sum of the amount present in the star at birth and 
that synthesized by the star. This was properly accounted for in our calculations.}
for massive stars and Marigo (2001) for low and intermediate mass stars (LIMS). An 
advantage of using these two sets is that they are designed to be complementary. The 
massive star models account for the effects of convective overshooting and 
quiescent mass loss and cover a metallicity range from 0.0004 to 0.05. For 
metallicities below 0.0004, yields were assumed to be equal to their values at 
0.0004, i.e. yields were not extrapolated in metallicity. However, since N 
production is exclusively secondary in massive stars, yields below 0.0004 were 
scaled with metallicity. Since Portinari et al. did not publish Cl and Ar 
yields, we used the results of Woosley \& Weaver (1995). Marigo's computations 
for LIMS account for mass loss, third dredge-up, and hot bottom burning over a 
metallicity range of 0.004 to 0.019. Finally, in the case of binary star 
contributions to evolution of C, O, Si, S, Cl, Ar, and Fe through SNIa events, 
we employed the yields of Nomoto et al. (1997), using both the W7 and W70 models 
to account for metallicity effects.

\section{MODEL UNIQUENESS}

In this appendix we consider the issue of model uniqueness within the confines of our assumptions, and in particular whether more than one SFE-age combination can adequately reproduce the results reported in Table~\ref{singlemodels}. To answer this question, we performed a Monte Carlo simulation in which input values of SFE and age were selected randomly
from within ranges whose extremes matched those of the values inferred for our sample and given in Table~\ref{singlemodels}. In this fashion 1000 parameter pairs were constructed. One thousand chemical evolution models were then calculated, one for each parameter pair, and the predicted values of N/Si and Si/H were plotted along with the observed values in the N/Si--Si/H plane.

Next, within this plane we selected seven widely spaced DLAs from our sample and proceeded to compare our {\it derived} values of SFE and age reported in Table~\ref{singlemodels} with the those of nearby models generated by the Monte Carlo simulation. (Since our parameter ranges are restricted from the beginning, this is clearly not a foolproof method. However, it should at least tell us whether or not there are alternative parameter combinations to the ones in Table~\ref{singlemodels} within those ranges.)

The results are shown in Table~\ref{uniqueness}, where for each of the seven DLAs, we list our derived parameters from Table~\ref{singlemodels} in the first row followed by analogous values for several simulation points in subsequent rows.

The parameter values appear to agree closely for all objects except Q0930+28, where our derived age of 1.2~Gyr is more than a factor of two greater than the ages of the nearby points. We conclude, however, that in most cases the conditions inferred in the individual models above are likely to be unique.

%

\clearpage

\begin{deluxetable}{lccccccccccc}
\tablewidth{0pc}
\tablecaption{STATISTICAL SAMPLE: DLA Abundance Summary\label{tab:stat}}
\tabletypesize{\tiny}
\tablehead{\colhead{Quasar} &\colhead{$z_{abs}$} & \colhead{log \nhi} & 
\colhead{log N} & \colhead{log Si} &
\colhead{log S} & \colhead{log Fe} &
\colhead{log(N/Si)} & \colhead{log (N/S)} &
\colhead{log(Si/Fe)} & \colhead{Ref}}
\startdata
Q0000-2619&3.3901&$21.41^{+0.08}_{-0.08}$&$ 14.73 \pm 0.03$&$ 15.06 \pm 0.02$&$$&$ 14.75 \pm 0.03$&$-0.33$&$...$&$ 0.31$&$2,6,9,10$\\
PH957&2.3090&$21.37^{+0.08}_{-0.08}$&$ 15.03 \pm 0.02$&$> 14.72$&$ 15.11 \pm 0.02$&$ 14.97 \pm 0.04$&$-0.43^a$&$-0.08$&$...$&$1,6,11$\\
Q0151+0448&1.9343&$20.36^{+0.10}_{-0.10}$&$ 12.90 \pm 0.11$&$ 13.96 \pm 0.02$&$$&$$&$-1.06$&$...$&$...$&$24$\\
Q0201+11&3.3869&$21.26^{+0.10}_{-0.10}$&$ 15.33 \pm 0.11$&$> 14.91$&$ 15.21 \pm 0.11$&$ 15.35 \pm 0.05$&$-0.23^a$&$ 0.12$&$...$&$12$\\
Q0201+36&2.4628&$20.38^{+0.05}_{-0.05}$&$> 15.00$&$ 15.53 \pm 0.02$&$ 15.29 \pm 0.02$&$ 15.01 \pm 0.02$&$>-0.53$&$>-0.29$&$ 0.53$&$3,11,13$\\
CTQ418&2.4292&$20.68^{+0.07}_{-0.07}$&$< 13.64$&$ 14.32 \pm 0.03$&$ 13.97 \pm 0.03$&$ 13.91 \pm 0.05$&$<-0.68$&$<-0.33$&$ 0.41$&$23$\\
CTQ418&2.5138&$20.50^{+0.07}_{-0.07}$&$ 13.88 \pm 0.03$&$ 14.52 \pm 0.03$&$ 13.99 \pm 0.06$&$ 14.07 \pm 0.03$&$-0.64$&$-0.11$&$ 0.45$&$23$\\
Q0336-01&3.0621&$21.20^{+0.10}_{-0.10}$&$> 15.04$&$> 15.01$&$ 14.86 \pm 0.02$&$ 14.89 \pm 0.02$&$< 0.03$&$> 0.18$&$...$&$11,13$\\
Q0347-38&3.0247&$20.63^{+0.00}_{-0.00}$&$ 14.89 \pm 0.03$&$ 15.02 \pm 0.03$&$< 14.77$&$ 14.50 \pm 0.02$&$-0.13$&$> 0.12$&$ 0.52$&$6,11,13,16$\\
CTQ247&2.5505&$21.13^{+0.10}_{-0.10}$&$ 14.55 \pm 0.03$&$ 15.32 \pm 0.04$&$ 14.82 \pm 0.06$&$ 14.95 \pm 0.06$&$-0.77$&$-0.27$&$ 0.37$&$23$\\
CTQ247&2.5950&$21.09^{+0.10}_{-0.10}$&$ 15.07 \pm 0.02$&$ 15.59 \pm 0.03$&$ 15.19 \pm 0.05$&$ 15.15 \pm 0.02$&$-0.52$&$-0.12$&$ 0.44$&$23$\\
CTQ247&2.6215&$20.47^{+0.10}_{-0.10}$&$< 14.36$&$ 13.99 \pm 0.06$&$< 14.34$&$ 13.60 \pm 0.02$&$< 0.37$&$< 0.02$&$ 0.39$&$23$\\
Q0450-13&2.0666&$20.53^{+0.08}_{-0.08}$&$ 13.76 \pm 0.03$&$ 14.62 \pm 0.02$&$ 14.29 \pm 0.04$&$ 14.15 \pm 0.03$&$-0.86$&$-0.53$&$ 0.47$&$21$\\
Q0528-2505&2.1410&$20.95^{+0.05}_{-0.05}$&$ 14.58 \pm 0.08$&$ 15.26 \pm 0.04$&$ 14.83 \pm 0.04$&$ 14.94 \pm 0.36$&$-0.68$&$-0.25$&$ 0.32$&$2$\\
HS0741+4741&3.0174&$20.48^{+0.10}_{-0.10}$&$ 13.98 \pm 0.02$&$ 14.35 \pm 0.02$&$ 14.01 \pm 0.02$&$ 14.05 \pm 0.02$&$-0.37$&$-0.02$&$ 0.30$&$11,13$\\
FJ0812+32&2.0668&$21.00^{+0.10}_{-0.10}$&$ 14.98 \pm 0.18$&$ 15.18 \pm 0.02$&$$&$ 14.89 \pm 0.02$&$-0.20$&$...$&$ 0.29$&$18,22,24$\\
FJ0812+32&2.6263&$21.35^{+0.10}_{-0.10}$&$ 16.00 \pm 0.34$&$ 15.98 \pm 0.05$&$ 15.63 \pm 0.08$&$ 15.09 \pm 0.02$&$ 0.02$&$ 0.37$&$ 0.89$&$18,22,24$\\
Q0841+12&2.3745&$20.95^{+0.09}_{-0.09}$&$ 14.61 \pm 0.03$&$ 15.24 \pm 0.02$&$ 14.77 \pm 0.03$&$$&$-0.63$&$-0.16$&$...$&$6,11$\\
Q0841+12&2.4762&$20.78^{+0.10}_{-0.10}$&$ 13.94 \pm 0.09$&$> 14.49$&$ 14.59 \pm 0.03$&$ 14.53 \pm 0.05$&$-1.00^a$&$-0.65$&$...$&$6,11$\\
J0900+4215&3.2458&$20.30^{+0.10}_{-0.10}$&$ 14.15 \pm 0.02$&$> 14.95$&$ 14.65 \pm 0.02$&$ 14.54 \pm 0.02$&$-0.85^a$&$-0.50$&$...$&$22,24$\\
Q0930+28&3.2353&$20.30^{+0.10}_{-0.10}$&$ 13.75 \pm 0.02$&$ 13.89 \pm 0.02$&$< 13.62$&$ 13.72 \pm 0.02$&$-0.14$&$> 0.13$&$ 0.17$&$13$\\
Q1008+36&2.7989&$20.70^{+0.05}_{-0.05}$&$ 13.30 \pm 0.04$&$ 14.46 \pm 0.02$&$$&$< 15.11$&$-1.15$&$...$&$...$&$11,13$\\
Q1055+46&3.3172&$20.34^{+0.10}_{-0.10}$&$< 14.09$&$ 14.25 \pm 0.11$&$$&$ 13.97 \pm 0.02$&$<-0.16$&$...$&$ 0.28$&$5$\\
BR1117-1329&3.3504&$20.84^{+0.10}_{-0.10}$&$< 14.53$&$ 15.13 \pm 0.08$&$$&$ 14.83 \pm 0.03$&$<-0.60$&$...$&$ 0.30$&$19$\\
Q1157+014&1.9440&$21.80^{+0.10}_{-0.10}$&$> 15.10$&$> 15.93$&$$&$ 15.49 \pm 0.04$&$<-0.83$&$...$&$...$&$8$\\
BR1202-07&4.3829&$20.60^{+0.14}_{-0.14}$&$ 13.81 \pm 0.11$&$ 14.35 \pm 0.02$&$$&$ 13.91 \pm 0.12$&$-0.54$&$...$&$ 0.44$&$2$\\
Q1210+17&1.8918&$20.60^{+0.10}_{-0.10}$&$ 14.71 \pm 0.11$&$ 15.29 \pm 0.02$&$ 14.96 \pm 0.02$&$ 14.96 \pm 0.06$&$-0.58$&$-0.25$&$ 0.33$&$11,21$\\
Q1223+17&2.4661&$21.50^{+0.10}_{-0.10}$&$ 14.83 \pm 0.18$&$ 15.47 \pm 0.02$&$$&$ 15.16 \pm 0.02$&$-0.64$&$...$&$ 0.31$&$7,11$\\
Q1232+08&2.3371&$20.90^{+0.10}_{-0.10}$&$ 14.63 \pm 0.09$&$ 15.18 \pm 0.09$&$ 14.83 \pm 0.11$&$ 14.68 \pm 0.09$&$-0.55$&$-0.20$&$ 0.50$&$8$\\
Q1331+17&1.7764&$21.14^{+0.08}_{-0.08}$&$ 15.20 \pm 0.11$&$ 15.28 \pm 0.02$&$$&$ 14.62 \pm 0.02$&$-0.08$&$...$&$ 0.67$&$6,11$\\
Q1409+095&2.4562&$20.54^{+0.10}_{-0.10}$&$< 13.19$&$ 14.08 \pm 0.02$&$$&$ 13.74 \pm 0.02$&$<-0.89$&$...$&$ 0.34$&$19$\\
Q1425+6039&2.8268&$20.30^{+0.04}_{-0.04}$&$ 14.71 \pm 0.02$&$> 15.02$&$$&$ 14.48 \pm 0.02$&$<-0.31$&$...$&$...$&$2,11,13,22$\\
J1435+5359&2.3427&$21.05^{+0.10}_{-0.10}$&$ 14.67 \pm 0.02$&$ 15.13 \pm 0.02$&$ 14.78 \pm 0.05$&$$&$-0.46$&$-0.11$&$...$&$24$\\
J1558-0031&2.7026&$20.67^{+0.05}_{-0.05}$&$ 14.46 \pm 0.02$&$ 14.24 \pm 0.02$&$ 14.07 \pm 0.02$&$ 14.11 \pm 0.03$&$ 0.22$&$ 0.39$&$ 0.13$&$20$\\
J1712+5755&2.2535&$20.60^{+0.10}_{-0.10}$&$ 13.89 \pm 0.03$&$ 14.92 \pm 0.06$&$$&$ 14.49 \pm 0.02$&$-1.03$&$...$&$ 0.43$&$24$\\
Q1759+75&2.6253&$20.76^{+0.01}_{-0.01}$&$ 15.11 \pm 0.02$&$ 15.53 \pm 0.02$&$ 15.24 \pm 0.02$&$ 15.08 \pm 0.02$&$-0.43$&$-0.13$&$ 0.46$&$6,11$\\
J2036-0553&2.2803&$21.20^{+0.15}_{-0.15}$&$ 14.87 \pm 0.03$&$ 15.04 \pm 0.05$&$$&$ 14.68 \pm 0.11$&$-0.17$&$...$&$ 0.36$&$24$\\
Q2206-19&2.0762&$20.43^{+0.06}_{-0.06}$&$ 12.58 \pm 0.05$&$ 13.68 \pm 0.04$&$$&$ 13.32 \pm 0.02$&$-1.10$&$...$&$ 0.36$&$4,6,11$\\
Q2230+02&1.8644&$20.85^{+0.08}_{-0.08}$&$ 15.02 \pm 0.03$&$ 15.65 \pm 0.02$&$ 15.29 \pm 0.03$&$ 15.19 \pm 0.02$&$-0.63$&$-0.27$&$ 0.47$&$6,11$\\
Q2231-002&2.0661&$20.56^{+0.10}_{-0.10}$&$< 15.02$&$ 15.24 \pm 0.02$&$ 15.10 \pm 0.18$&$ 14.66 \pm 0.07$&$<-0.22$&$<-0.08$&$ 0.58$&$2,6,11,17$\\
BR2237-0607&4.0803&$20.52^{+0.11}_{-0.11}$&$< 14.28$&$ 14.21 \pm 0.02$&$$&$ 13.88 \pm 0.12$&$< 0.07$&$...$&$ 0.33$&$2$\\
HE2243-6031&2.3300&$20.67^{+0.02}_{-0.02}$&$ 14.88 \pm 0.03$&$ 15.36 \pm 0.03$&$ 15.02 \pm 0.03$&$ 14.92 \pm 0.03$&$-0.48$&$-0.14$&$ 0.44$&$14$\\
Q2342+34&2.9082&$21.10^{+0.10}_{-0.10}$&$ 14.92 \pm 0.04$&$ 15.62 \pm 0.02$&$ 15.19 \pm 0.02$&$ 15.02 \pm 0.06$&$-0.70$&$-0.28$&$ 0.60$&$18,22$\\
Q2344+12&2.5379&$20.36^{+0.10}_{-0.10}$&$ 13.79 \pm 0.03$&$ 14.18 \pm 0.02$&$< 14.19$&$ 14.04 \pm 0.03$&$-0.39$&$>-0.40$&$ 0.14$&$2,11,13$\\
Q2348-14&2.2794&$20.56^{+0.08}_{-0.08}$&$ 13.35 \pm 0.06$&$ 14.20 \pm 0.02$&$ 13.70 \pm 0.13$&$ 13.82 \pm 0.02$&$-0.85$&$-0.35$&$ 0.39$&$6,11,13$\\
\enddata
\tablenotetext{a}{Calculated from log(N/S) and offset by the solar relative
abundance, $-0.35$dex.}
\tablecomments{Note that none of the limits reported take into account the
uncertainty in the \nhi\ value.}
\tablerefs{
1: Wolfe et al. (1994);
2: Lu et al. (1996);
3: Prochaska \& Wolfe (1996);
4: Prochaska \& Wolfe (1997);
5: Lu, Sargent, \& Barlow (1999);
6: Prochaska \& Wolfe (1999);
7: Prochaska \& Wolfe (2000);
8: Petitjean, Srianand, \& Ledoux (2000);
9: Molaro et al. (2000);
10: Molaro et al. (2001);
11: Prochaska et al. (2001);
12: Ellison et al. (2001);
13: Prochaska et al. (2002);
14: Lopez et al. (2002);
15: Levshakov et al. (2002);
16: Lopez \& Ellison (2003);
17: Dessauges-Zavadsky et al. (2004);
18: Prochaska et al. (2003);
19: Ledoux, Petitjean, \& Srianand (2003);
20: O'Meara et al. (2006);
21: Dessauges-Zavaksky et al. (2006);
22: Prochaska et al. (2007);
23: Lopez et al., in prep;
24: Dessauges-Zavadsky et al., in prep}
 
\end{deluxetable}


\begin{deluxetable}{lccccccccccc}
\tablewidth{0pc}
\setlength{\tabcolsep}{0.03in}
\tablecaption{CHEMICAL EVOLUTION MODELS\label{singlemodels}}
\tabletypesize{\tiny}
\rotate
\tablehead{
\colhead{Quasar} &
\colhead{SFE (Gyr$^{-1}$)} & 
\colhead{Age (Gyr)} & 
\colhead{log(N/H)\tablenotemark{1}} & 
\colhead{log(Si/H)\tablenotemark{1}} &
\colhead{log(S/H)\tablenotemark{1}} & 
\colhead{log(Fe/H)\tablenotemark{1}} &
\colhead{log(O/H)\tablenotemark{1}} &
\colhead{log(C/H)\tablenotemark{2}} &
\colhead{log(N/Si)\tablenotemark{1}} & 
\colhead{log(Si/Fe)\tablenotemark{1}} &
\colhead{log(C/Si)\tablenotemark{2}}
}
\startdata

Q0000-2619 &0.05 &0.50 &-6.68/-6.71 &-6.35/-6.38 &\nodata/-6.72 &-6.66/-6.70 &-4.99/-4.98 &-6.17 &-0.33/-0.33 &+0.31/+0.31 &+0.21 \\
Q0151+0448 &0.13 &0.20 &-7.46/-7.49 &-6.40/-6.41 &\nodata/-6.75 &\nodata/-6.76 &\nodata/-5.00 &-6.21 &-1.06/-1.08 &\nodata/+0.35 &+0.20 \\
CTQ418 &0.25 &0.28 &-6.62/-6.62 &-5.98/-5.97 &-6.51/-6.31 &-6.43/-6.30 &\nodata/-4.56 &-5.78 &-0.64/-0.65 &+0.45/+0.34 &+0.19 \\
Q0347-38 &0.18 &0.80 &-5.74/-5.76 &-5.61/-5.62 &$<$-5.86/-5.96 &-6.13/-5.91 &-4.67/-4.23 &-5.37 &-0.13/-0.14 &+0.52/+0.29 &+0.25 \\
CTQ247 &0.40 &0.25 &-6.58/-6.55 &-5.81/-5.82 &-6.31/-6.16 &-6.18/-6.16 &\nodata/-4.41 &-5.63 &-0.77/-0.73 &+0.37/+0.34 &+0.19 \\
CTQ247 &0.90 &0.27 &-6.02/-6.01 &-5.50/-5.45 &-5.90/-5.79 &-5.94/-5.78 &\nodata/-4.04 &-5.26 &-0.52/-0.56 &+0.44/+0.33 &+0.19 \\
Q0450-13 &0.40 &0.22 &-6.77/-6.74 &-5.91/-5.88 &-6.24/-6.22 &-6.38/-6.23 &-5.12/-4.70 &-5.69 &-0.86/-0.86 &+0.47/+0.34 &+0.19 \\
Q0528-2505 &0.50 &0.26 &-6.37/-6.37 &-5.69/-5.71 &-6.12/-6.05 &-6.01/-6.05 &\nodata/-4.30 &-5.52 &-0.68/-0.66 &+0.32/+0.34 &+0.19 \\
HS0741+4741 &0.10 &0.45 &-6.51/-6.51 &-6.13/-6.14 &-6.47/-6.47 &-6.43/-6.45 &-4.76/-4.73 &-5.94 &-0.38/-0.37 &+0.30/+0.32 &+0.20 \\
FJ0812+32 &0.12 &0.70 &-6.02/-6.04 &-5.82/-5.85 &\nodata/-6.19 &-6.11/-6.15 &\nodata/-4.46 &-5.61 &-0.20/-0.19 &+0.29/+0.29 &+0.24 \\
FJ0812+32 &0.10 &2.00 &-5.35/-5.46 &-5.37/-5.44 &-5.72/-5.78 &-6.26/-5.68 &-3.86/-4.07 &-5.05 &+0.02/-0.02 &+0.89/+0.23 &+0.39 \\
Q0841+12 &0.50 &0.28 &-6.34/-6.27 &-5.71/-5.67 &-6.18/-6.01 &\nodata/-6.01 &\nodata/-4.27 &-5.48 &-0.63/-0.60 &\nodata/+0.33 &+0.19 \\
Q0930+28 &0.02 &1.20 &-6.55/-6.48 &-6.41/-6.37 &$<$-6.68/-6.70 &-6.58/-6.64 &-5.01/-4.98 &-6.06 &-0.14/-0.11 &+0.17/+0.27 &+0.31 \\
Q1008+36 &0.20 &0.19 &-7.40/-7.37 &-6.24/-6.25 &\nodata/-6.59 &$<$-5.59/-6.60 &-4.93/-4.84 &-6.05 &-1.16/-1.12 &$>$-0.65/+0.35 &+0.20 \\
BR1202-07 &0.10 &0.35 &-6.79/-6.76 &-6.25/-6.25 &\nodata/-6.59 &-6.69/-6.58 &\nodata/-4.85 &-6.05 &-0.54/-0.51 &+0.44/+0.33 &+0.20 \\
Q1210+17 &1.30 &0.25 &-5.89/-5.89 &-5.31/-5.33 &-5.64/-5.67 &-5.64/-5.66 &\nodata/-3.93 &-5.15 &-0.58/-0.56 &+0.33/+0.33 &+0.18 \\
Q1223+17 &0.23 &0.28 &-6.67/-6.66 &-6.03/-6.00 &\nodata/-6.34 &-6.34/-6.34 &-6.02/-4.59 &-5.81 &-0.64/-0.66 &+0.31/+0.34 &+0.19 \\
Q1232+08 &0.40 &0.30 &-6.27/-6.30 &-5.72/-5.74 &-6.07/-6.07 &-6.22/-6.07 &\nodata/-4.33 &-5.55 &-0.55/-0.56 &+0.50/+0.33 &+0.19 \\
Q1331+17 &0.06 &1.20 &-5.94/-6.00 &-5.86/-5.90 &\nodata/-6.24 &-6.52/-6.17 &\nodata/-4.51 &-5.59 &-0.08/-0.10 &+0.66/+0.27 &+0.31 \\
J1435+5359 &0.20 &0.40 &-6.38/-6.31 &-5.92/-5.90 &-6.27/-6.23 &\nodata/-6.22 &\nodata/-4.49 &-5.70 &-0.46/-0.41 &\nodata/+0.32 &+0.20 \\
J1558-0031\tablenotemark{3} &0.01 &2.20 &-6.21/-6.33 &-6.43/-6.43 &-6.60/-6.72 &-6.56/-6.61 &-4.81/-5.01 &-6.02 &+0.22/+0.10 &+0.13/+0.24 &+0.41 \\
J1712+5755 &0.80 &0.18 &-6.71/-6.69 &-5.68/-5.69 &\nodata/-6.03 &-6.11/-6.04 &\nodata/-4.27 &-5.50 &-1.03/-1.00 &+0.43/+0.35 &+0.19 \\
Q1759+75 &1.50 &0.28 &-5.65/-5.67 &-5.23/-5.22 &-5.52/-5.57 &-5.68/-5.55 &-4.50/-3.82 &-5.04 &-0.42/-0.45 &+0.45/+0.32 &+0.18 \\
J2036-0553 &0.05 &0.80 &-6.33/-6.35 &-6.16/-6.17 &\nodata/-6.50 &-6.52/-6.45 &\nodata/-4.77 &-5.92 &-0.17/-0.18 &+0.36/+0.29 &+0.25 \\
Q2206-19 &0.06 &0.20 &-7.85/-7.86 &-6.75/-6.75 &\nodata/-7.09 &-7.11/-7.10 &-5.89/-5.33 &-6.55 &-1.10/-1.11 &+0.36/+0.35 &+0.20 \\
Q2230+02 &2.00 &0.20 &-5.83/-5.93 &-5.20/-5.26 &-5.56/-5.60 &-5.66/-5.60 &\nodata/-3.85 &-5.08 &-0.63/-0.67 &+0.46/+0.34 &+0.18 \\
HE2243-6031 &1.30 &0.28 &-5.79/-5.75 &-5.31/-5.28 &-5.65/-5.62 &-5.75/-5.61 &\nodata/-3.88 &-5.10 &-0.48/-0.47 &+0.44/+0.33 &+0.18 \\
Q2342+34 &1.00 &0.22 &-6.21/-6.23 &-5.48/-5.50 &-5.91/-5.84 &-6.08/-5.84 &-5.57/-4.09 &-5.31 &-0.73/-0.73 &+0.60/+0.34 &+0.19 \\
Q2344+12 &0.10 &0.43 &-6.57/-6.55 &-6.18/-6.16 &$<$-6.17/-6.49 &-6.32/-6.48 &-5.33/-4.75 &-5.96 &-0.39/-0.39 &+0.14/+0.32 &+0.20 \\
Q2348-14 &0.13 &0.23 &-7.21/-7.23 &-6.36/-6.34 &-6.86/-6.68 &-6.74/-6.69 &-5.71/-4.93 &-6.15 &-0.85/-0.89 &+0.38/+0.34 &+0.19 \\
Average Uncertainty\tablenotemark{4} & \nodata & \nodata & 0.11 & 0.089 & 0.090 & 0.10 & 0.27 & \nodata & 0.073 & 0.056 & \nodata \\
Sun\tablenotemark{5} & \nodata & \nodata & -4.22 & -4.49 & -4.84 & -4.55 & -3.34 & -3.61 & +0.27 & +0.06 & +0.88 

\enddata

\tablenotetext{1}{The observed and model values are expressed as observed/model, with typical observational uncertainties of $\le$0.1~dex.}
\tablenotetext{2}{Model results only; No observed values are available.}
\tablenotetext{3}{N yields increased.}
\tablenotetext{4}{Square root of the quadrature sum of the logarithmic uncertainties in the observed numerator and denominator values.}
\tablenotetext{5}{Asplund et al. 2005}

\end{deluxetable}


\begin{deluxetable}{lccccccccccc}
\tablewidth{0pc}
\setlength{\tabcolsep}{0.03in}
\tablecaption{CHEMICAL EVOLUTION MODELS: ALTERED IMF\label{altimf}}
\tabletypesize{\scriptsize}
\rotate
\tablehead{
\colhead{Quasar} &
\colhead{SFE (Gyr$^{-1}$)} & 
\colhead{Age (Gyr)} & 
\colhead{log(N/H)\tablenotemark{1}} & 
\colhead{log(Si/H)\tablenotemark{1}} &
\colhead{log(S/H)\tablenotemark{1}} & 
\colhead{log(Fe/H)\tablenotemark{1}} &
\colhead{log(O/H)\tablenotemark{1}} &
\colhead{log(C/H)\tablenotemark{2}} &
\colhead{log(N/Si)\tablenotemark{1}} & 
\colhead{log(Si/Fe)\tablenotemark{1}} &
\colhead{log(C/Si)\tablenotemark{2}} 
}
\startdata
\cutinhead{Truncated IMF\tablenotemark{3}}
Q0151+0448 &0.006 &3.00 &-7.46/-7.49 &-6.40/-6.46 &\nodata/-6.79 &\nodata/-6.68 &\nodata/-5.08 &-6.06 &-1.06/-1.03 &\nodata/+0.22 &+0.40 \\
Q1008+36 &0.10 &0.40 &-7.40/-7.32 &-6.24/-6.19 &\nodata/-6.53 &$<$-5.59/-6.51 &-4.93/-4.79 &-5.78 &-1.16/-1.13 &-0.65/+0.32 &+0.41 \\
Q2206-19 &0.003 &3.00 &-7.85/-7.81 &-6.75/-6.76 &\nodata/-7.09 &-7.11/-6.98 &-5.89/-5.38 &-6.36 &-1.10/-1.05 &+0.36/+0.22 &+0.40  \\
Q2348-14 &0.009 & 2.70 &-7.21/-7.13 &-6.36/-6.34 &-6.86/-6.67 &-6.74/-6.56 &-5.71/-4.96 &-5.94 &-0.85/-0.79 &+0.38/+0.22 &+0.40 \\
\cutinhead{Flat IMF\tablenotemark{4}}
Q0151+0448 &0.05 &0.20 &-7.46/-7.49 &-6.40/-6.42 &\nodata/-6.76 &\nodata/-6.88 &\nodata/-4.54 &-5.64 &-1.06/-1.07 &\nodata/+0.46 &+0.78  \\
Q2206-19 &0.01 &0.40 &-7.85/-7.82 &-6.75/-6.78 &\nodata/-7.13 &-7.11/-7.22 &-5.89/-4.91 &-6.01 &-1.10/-1.04 &+0.36/+0.44 &+0.77  \\
Q2348-14 &0.02 &0.51 &-7.21/-7.17 &-6.36/-6.37 &-6.86/-6.71 &-6.74/-6.80 &-5.71/-4.51 &-5.60 &-0.85/-0.80 &+0.38/+0.43 &+0.77  \\
\cutinhead{Normal IMF}
Q0151+0448 &0.10 &0.20 &-7.46/-7.49 &-6.40/-6.41 &\nodata/-6.75 &\nodata/-6.76 &\nodata/-5.00 &-6.21 &-1.06/-1.08 &\nodata/+0.35 &+0.20 \\
Q1008+36 &0.20 &0.19 &-7.40/-7.37 &-6.24/-6.25 &\nodata/-6.59 &$<$-5.59/-6.60 &-4.93/-4.84 &-6.05 &-1.16/-1.12 &-0.65/+0.35 &+0.20 \\
Q2206-19 &0.06 &0.20 &-7.85/-7.86 &-6.75/-6.75 &\nodata/-7.09 &-7.11/-7.10 &-5.89/-5.33 &-6.55 &-1.10/-1.11 &+0.36/+0.35 &+0.20 \\
Q2348-14 &0.13 &0.23 &-7.21/-7.23 &-6.36/-6.34 &-6.86/-6.68 &-6.74/-6.69 &-5.71/-4.93 &-6.15 &-0.85/-0.89 &+0.38/+0.34 &+0.19 \\

\enddata

\tablenotetext{1}{The observed and model values are expressed as observed/model.}
\tablenotetext{2}{Model results only; No observed values are available.}
\tablenotetext{3}{IMF is truncated below 4.80~M$_{\odot}$ for all but Q2348-14, where 4.65~M$_{\odot}$ is used.}
\tablenotetext{4}{The IMF slope is $\alpha$=0.10}

\end{deluxetable}


\begin{deluxetable}{lcc}
\tablewidth{0pc}
\setlength{\tabcolsep}{0.03in}
\tablecaption{MODEL UNIQUENESS TEST RESULTS\label{uniqueness}}
\tabletypesize{\tiny}
\tablehead{
\colhead{Data Source\tablenotemark{1}} &
\colhead{SFE (Gyr$^{-1}$)} & 
\colhead{Age (Gyr)} 
}
\startdata
\cutinhead{Q2206}
Model & 0.060 & 0.20 \\
S1 & 0.046 & 0.23 \\
S2 & 0.039 & 0.28 \\
S3 & 0.065 & 0.21 \\
S4 & 0.039 & 0.32 \\
S5 & 0.070 & 0.20 \\
\cutinhead{Q0151}
Model & 0.13 & 0.20 \\
S1 & 0.14 & 0.20 \\
S2 & 0.10 & 0.25 \\
S3 & 0.094 & 0.26 \\
S4 & 0.13 & 0.19 \\
\cutinhead{Q0930}
Model & 0.02 & 1.2 \\
S1 & 0.050 & 0.49 \\
S2 & 0.058 & 0.42 \\
S3 & 0.05 & 0.49 \\
\cutinhead{Q1759}
Model & 1.50 & 0.28 \\
S1 & 1.36 & 0.30 \\
S2 & 1.33 & 0.31 \\
S3 & 1.32 & 0.31 \\
S4 & 1.28 & 0.32 \\
S5 & 1.34 & 0.30 \\
\cutinhead{Q0450}
Model & 0.40 & 0.22 \\
S1 & 0.39 & 0.21 \\
S2 & 0.38 & 0.21 \\
S3 & 0.35 & 0.23 \\
S4 & 0.44 & 0.20 \\
S5 & 0.34 & 0.23 \\
\cutinhead{CTQ247 (z$_{abs}$=2.595)}
Model & 0.90 & 0.27 \\
S1 & 0.72 & 0.30 \\
S2 & 0.78 & 0.28 \\
S3 & 0.81 & 0.27 \\
S4 & 0.78 & 0.27 \\
S5 & 0.66 & 0.31 \\
\cutinhead{Q1210}
Model & 1.30 & 0.25 \\
S1 & 1.39 & 0.25 \\
S2 & 1.39 & 0.25 \\
S3 & 1.33 & 0.26 \\
S4 & 1.28 & 0.27 \\
S5 & 1.40 & 0.25 \\

\enddata

\tablenotetext{1}{Values of SFE and age pertaining to the model in Table~\ref{singlemodels} are listed in the first line for each object. Each subsequent line provides values for several simulation points.}

\end{deluxetable}


\clearpage

\end{document}